\documentclass[prl,aps,twocolumn,showpacs]{revtex4}
\usepackage{amsmath}
\usepackage{amsfonts}
\usepackage{amssymb}
\usepackage{graphicx}
\usepackage{color}

\renewcommand{\S}{\mathcal{S}}

\newcommand{\beq}{\begin{equation}}
\newcommand{\eeq}{\end{equation}}

\begin{document}

\title{Elements of Coevolution in Biological Sequences}

\author{Olivier Rivoire} 

\address{CNRS/UJF-Grenoble 1, LIPhy UMR 5588, Grenoble, F-38402, France}

\begin{abstract}
Studies of coevolution of amino acids within and between proteins have revealed two types of coevolving units: coevolving contacts, which are pairs of amino acids distant along the sequence but in contact in the three-dimensional structure, and sectors, which are larger groups of structurally connected amino acids that underlie the biochemical properties of proteins. By reconciling two approaches for analyzing correlations in multiple sequence alignments, we link these two findings together and with coevolving units of intermediate size, called `sectons', which are shown to provide additional information. By extending the analysis to the co-occurrence of orthologous genes in bacterial genomes, we also show that the methods and results are general and relevant beyond protein structures.
\end{abstract}

% 02.50.Sk multivariate analysis
% 87.14.E- proteins
% 87.15.Qt sequence analysis
% 87.18.Wd genomics

\maketitle

The structural and functional properties of proteins emerge from interactions between their amino acids. During evolution, these interactions constrain the substitutions of amino acids that may happen. Sequences resulting from multiple independent evolutionary trajectories reflect these constraints, and therefore contain information about the organization of interactions within proteins. Such sequences are now made available by DNA sequencing technology, which provides thousands of protein sequences that have diverged independently and under similar selective pressures from a common ancestral sequence.

These protein sequences are commonly collected into multi-sequence alignments on the basis of their sequence similarity. An alignment is formally an $M\times L$ array $X$, where $X_{si}$ indicates which of the $A=20$ natural amino acids is present at position $i$ in sequence $s$; some positions contain a gap, inserted to ensure an optimal alignment and represented as a 21st amino acid. Typical numbers are $M\sim 10^2$-$10^4$ for the number of sequences and $L\sim 10^2$-$10^3$ for the length of the alignment. 

The pattern of functional couplings between amino acids may be inferred from the statistical correlations between pairs of positions in the alignment. Analyses of these correlations are complicated by several factors: (i) proteins are gathered in an alignment based on sequence similarity, with no guarantee to have been subject to common selective constraints; (ii) sequences are not sampled independently during evolution, but through a branching process, which introduces a sampling bias;  (iii) the information content of the alignment, $\sim ML\log_2 A\sim 10^5$-$10^7$ bits, is small compared to the number  $\sim A^2 L^2/2\sim 10^6$-$10^8$ of continuous parameters defining the correlations between every pair of amino acids, which implies a severe under-sampling; (iv) two positions may be correlated while not directly interacting, reflecting a fundamental difference between interactions and correlations. 

Standard statistical analyses identify the observed samples to an asymptotically large number of independently and identically distributed random variables. Points (i), (ii) and (iii) violate each of these assumptions, while point (iv) suggests that, even in the absence of bias, further processing is required to infer interactions from correlations.

Many approaches have been proposed to tackle these challenges~\cite{Horner:2007if}. Recently, two methods have been developed, each rooted in a different concept of statistical mechanics. In an extension of an approach called Statistical Coupling Analysis (SCA)~\cite{Lockless:1999uf}, an application of concepts from random matrix theory~\cite{Plerou:2002im} to address (iii) has revealed collective modes of coevolution named sectors~\cite{Halabi:2009jca}. A protein sector is a group of structurally contacting positions, and experiments indicate that each sector controls independently a biochemical property of the protein~\cite{Halabi:2009jca}. In a different approach called Direct Coupling Analysis (DCA)~\cite{Weigt:2009we}, the problem (iv) of inferring interactions from correlations was formulated and solved as a problem of inverse statistical mechanics, leading to the inference of a large number of pairs of positions in contact in the folded structure~\cite{Morcos:2011jg}. 

The two approaches, SCA and DCA, differ in their principles as well as in their results. Using the  Pfam alignment for the trypsin family~\cite{Punta:2011ko} as an illustrative example, we show here how they can be connected at different levels. Specifically, we show that: (1) their respective measures of coevolution rely on distinct parts of the spectrum of a same covariance matrix; (2) a parallel analysis of the two measures of coevolution reveals different types of coevolving units - previously identified sectors and smaller units, which we call `sectons'; (3) these coevolving units, and the contacting pairs from DCA, stem from different aspects of the data but are interrelated, with sectons and contacting pairs respecting the overarching decomposition into independent sectors.

Given a multiple sequence alignment, SCA and DCA use as input the same basic statistical quantities: the frequency $f_i^{a}$ of amino acid $a$ at position $i$, and the joint frequency $f_{ij}^{ab}$ of the pair of amino acids $a,b$ at the pair of positions $i,j$. Prior to defining these frequencies, some steps must be taken to clean the alignment from positions with excessive gaps and mitigate the effects of (i) and (ii) by weighting differentially the contributions of the various sequences. These steps are straightforward but essential, and may be common for both approaches (all details are provided as Supplementary Material~\cite{SI}).

The frequencies $f_{ij}^{ab}$ and $f_i^{a}$ define a covariance matrix $C_{ij}^{ab}=f_{ij}^{ab}-f_i^{a}f_i^{b}$. SCA combines this matrix with a measure of amino acid conservation to define a matrix of conserved correlations $\mathcal{C}_{ij}$, while DCA relies on the inverse $J=-\bar C^{-1}$ of a regularized variant of $C_{ij}^{ab}$ (see below) to define a matrix of direct information $\mathcal{D}_{ij}$~\cite{SI}. Inspired by previous applications of random matrix theory to the study of covariance matrices~\cite{Laloux:1999wr,Plerou:2002im}, we analyze here these two matrices by a common method: (1) we compute the eigenvectors associated with the top $k_{\rm top}$ eigenvalues; (2) we rotate these eigenvectors into maximally independent components, $V^{(1)},\dots V^{(k_{\rm top})}$, using independent component analysis (ICA)~\cite{Bell:1995vn}; (3) we define coevolving units as sets of positions making largest contributions to a component, $\mathcal{S}_k=\{i : V^{(k)}_i>\epsilon\}$. The analysis involves two cutoffs: the number $k_{\rm top}$ of modes that is retained, and a threshold $\epsilon>0$ of significance for the contribution of positions to the components. The results, however, will be shown to be insensitive to the exact values of these cutoffs.

\begin{figure}[t]
\begin{center}
\includegraphics[width=.9\linewidth]{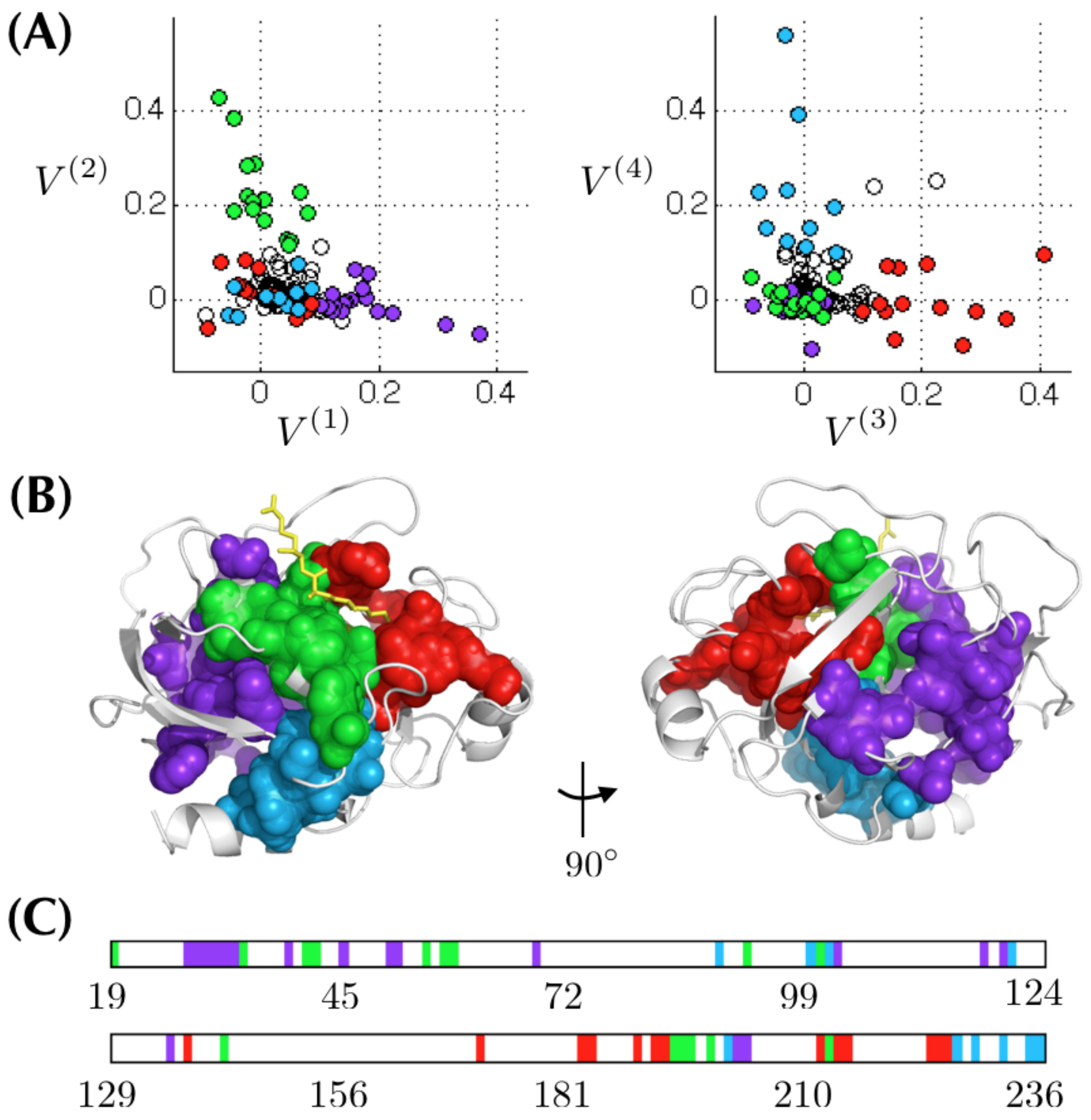}
\caption{Protein sectors in the trypsin family, as inferred from the Pfam alignment PF00089~\cite{Punta:2011ko} -- {\bf (A)} Projections of the positions $i$ along the vectors $V^{(k)}$ obtained by rotating by ICA the top $k_{\rm top}=4$  eigenvectors of the SCA matrix $\mathcal{C}_{ij}$~\cite{Lockless:1999uf}: each dot corresponds to a position $i$, with coordinates $(V^{(1)}_i,V^{(2)}_i)$ in the first graph, and $(V^{(3)}_i,V^{(4)}_i)$ in the second. Sector $k$ is defined by the positions $i$ with $V_i^{(k)}>\epsilon$ and $V_i^{(\ell)}<\epsilon$ for $\ell\neq k$, with $\epsilon=0.1$. The positions of each sector are represented with a different color: purple ($k=1$), green ($k=2$), red ($k=3$), cyan ($k=4$).  {\bf (B)} Location of the sectors on a three-dimensional structure of trypsin~\cite{Pasternak:1999bxa}. {\bf (C)} Location of the sectors along the sequence (cut in two for readability), with non-sector positions in white (numbering system of bovine chymotrypsin).}
\label{fig:FigSectors}
\end{center}
\end{figure}

For the SCA matrix $\mathcal{C}_{ij}$~\cite{Lockless:1999uf}, this analysis leads to coevolving units called protein sectors~\cite{Halabi:2009jca}. They are represented in Fig.~\ref{fig:FigSectors} for the alignment of the trypsin family, using $k_{\rm top}=4$ and $\epsilon=0.1$ (for simplicity, these sectors do not include the positions $i$ with $V^{(k)}_i>\epsilon$ for multiple $k$; Fig.~S1). Each sector forms a contacting group of positions on the three-dimensional structure, despite not necessarily consisting of consecutive positions along the sequence. Sectors have no sharp boundaries, but are typically organized into an onion-like hierarchy, with the core of sector $k$ consisting of  positions $i$ with largest $V^{(k)}_i$, and layers associated with decreasing values of $V^{(k)}_i$, as revealed by varying $\epsilon$ and $k_{\rm top}$ (Figs.~S2-6). Three sectors were previously inferred for the same protein family using an alignment about 10 times smaller~\cite{Halabi:2009jca}: two, the same green and red sectors, correspond to enzymatic activity and specificity respectively (Tab.~SI, Fig.~S7); the third one, which had the peculiarity of a disconnected core, and which correlated experimentally with stability, is now partly spread over two new sectors, whose functional role remains to be characterized.

\begin{figure}[t]
\begin{center}
\includegraphics[width=\linewidth]{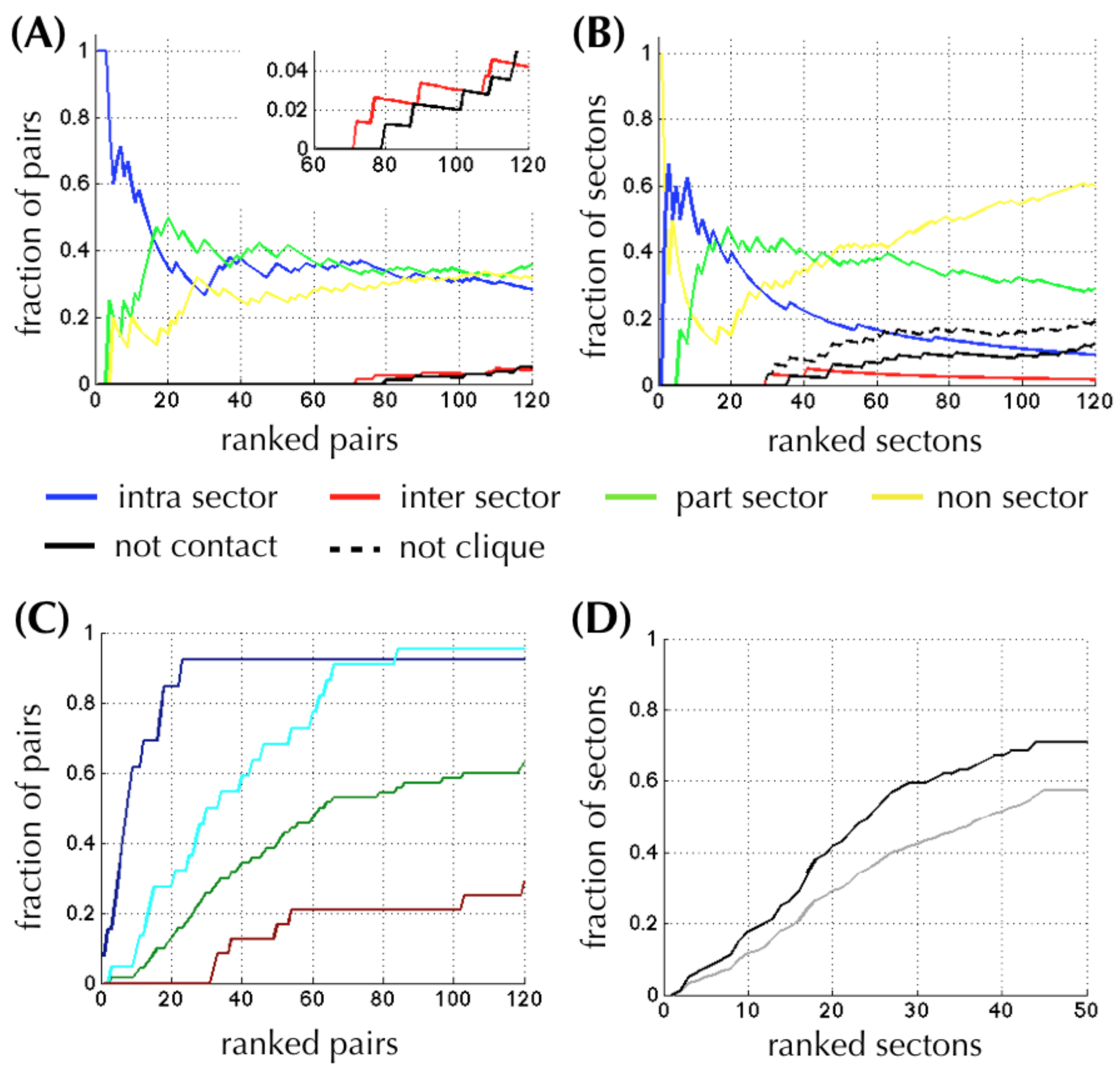}
\caption{Relations between top pairs of $\mathcal{\tilde D}_{ij}$, sectors and sectons -- {\bf (A)} Fraction of top pairs $ij$ of $\mathcal{\tilde D}_{ij}$, ranked by decreasing value of $\mathcal{\tilde D}_{ij}$, that are within a sector (blue curve), across two sectors (red), partly in a sector (green) and outside sectors (yellow). The fraction of pairs not in contact (black) becomes non zero at rank 80 (zoom in inset). {\bf (B)} Similar to (A), but for sectons instead of top pairs, and with an extra curve (dotted line) for the fraction of sectons that are not cliques, i.e., with two positions not directly in physical contact, but possibly contacting through other positions in the secton. As top pairs of $\mathcal{\tilde D}_{ij}$, sectons respect the decomposition into sectors. {\bf (C)} Fraction of contacting pairs within sectons of size 2 (blue) or size $\geq 3$ (green) that are top pairs of $\mathcal{\tilde D}_{ij}$, for the top 35 sectons that are structurally connected. Contacts in sectons of size $\geq 3$ can be partitioned into contacts associated with the 2 positions contributing most to the secton (cyan), which are nearly all top pairs of $\mathcal{\tilde D}_{ij}$, and other contacts (red), of which only $\sim 20\%$ are top pairs of $\mathcal{\tilde D}_{ij}$. {\bf (D)} Fraction of the top 79 (black) or 120 (gray) pairs of $\mathcal{\tilde D}_{ij}$ contained in a secton: $\sim 30\%$ of these top pairs are not in a secton. }
\label{fig:dcastat}
\end{center}
\end{figure}

DCA leads to a matrix $\mathcal{D}_{ij}$ of direct information, previously analyzed by ranking its entries~\cite{Weigt:2009we}: in a number of protein families, these top entries have been shown to consist of pairs of positions in physical contact in the three-dimensional structure~\cite{Morcos:2011jg} (contacts are defined here by a distance $< 8$ \AA). Most of these top pairs are however consecutive along the protein chain, due to the presence of stretches of gaps in the alignment. To discard these trivial contacts, we consider here a truncated matrix $\mathcal{\tilde D}_{ij}$, where $\mathcal{\tilde D}_{ij}=\mathcal{D}_{ij}$ if $|i-j|>\Delta$, and 0 otherwise, with $\Delta=5$ (other values give consistent results; Figs.~S8-9). For the trypsin alignment that serves here as illustration, the top 79 entries of this matrix are found to be in physical contact (Fig.~\ref{fig:dcastat}(A), Tab.~SII). The same figure shows that these contacts are not unrelated to sectors, but respect the decomposition into independent sectors, with top pairs of $\mathcal{\tilde D}_{ij}$ found within sectors, outside sectors, or at the edge of sectors, but almost never intersecting two sectors.

\begin{figure}[t]
\begin{center}
\includegraphics[width=\linewidth]{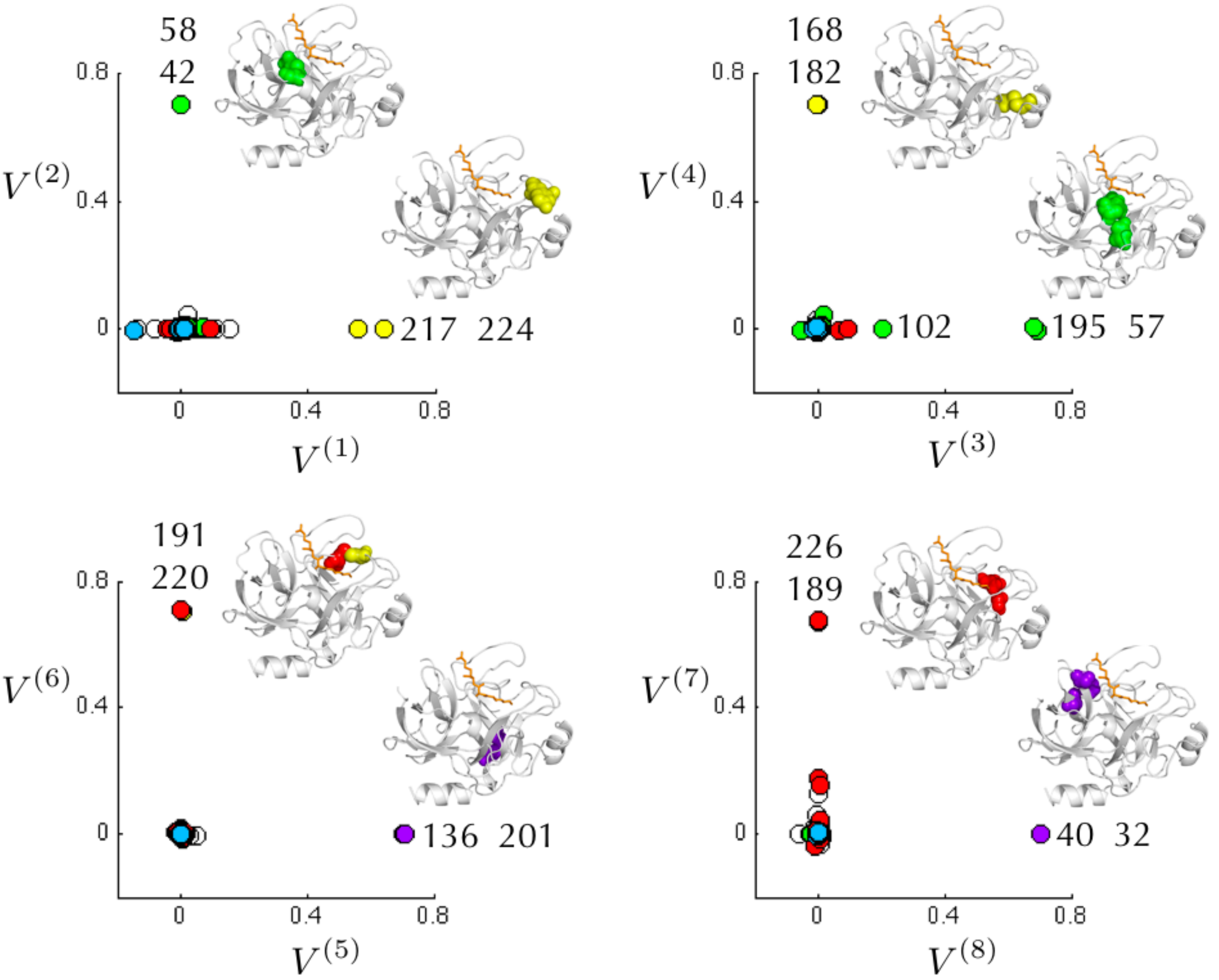}
\caption{Top protein sectons in the trypsin family -- Each graph is a projection of the positions along $(V^{(k)}, V^{(k+1)})$, the components of order $k$ and $k+1$ obtained by rotating by ICA the top eigenvectors of the truncated matrix of direct information $\mathcal{\tilde D}_{ij}$. Sectons are defined by $s_k=\{i: V_i^{(k)}>\epsilon\}$, with $\epsilon=0.2$. The labeling of positions follows the numbering system of bovine chymotrypsin (in several instances positions appear as superimposed) and the colors reflect the sectors as in Fig.~\ref{fig:FigSectors}, with yellow for non-sector positions. The location of the sectons on the three-dimensional structure is also indicated (more sectons are shown in Fig.~S10). Sectons $s_2,s_4,s_5,s_6$ are disulfide bonds, and $s_3$ is the catalytic triad. }
\label{fig:protsectons}
\end{center}
\end{figure}

Instead of considering the top elements of $\mathcal{\tilde D}_{ij}$, we also analyze here its spectral properties, following the method used to infer sectors from the SCA matrix $\mathcal{C}_{ij}$. This analysis leads to a large number ($\sim 100$) of independent components, each localized on a small group of 2 to 4 positions, which we call protein `sectons' (Tab.~SIII, Fig.~S10). Fig.~\ref{fig:protsectons} shows the first 8 sectons, using $k_{\rm top}=120$ and $\epsilon=0.2$, but similar results are obtained for a range of values of $k_{\rm top}$ and $\epsilon$ (Figs.~S11-12). As indicated in Fig.~\ref{fig:dcastat}(B), the first 35 sectons are structurally connected. Out of these 35 sectons, 13 have size 2, 20 size 3 and 2 size 4 (Fig.~S13). Fig.~\ref{fig:dcastat}(C) shows that sectons of size 2 are top pairs of $\mathcal{\tilde D}_{ij}$ (for technical reasons, an exception is the first secton; see Fig.~S12), but sectons of size $\geq 3$ include contacting pairs that are not top pairs of $\mathcal{\tilde D}_{ij}$. Reciprocally, Fig.~\ref{fig:dcastat}(D) shows that $\sim 60\%$ of the top pairs of $\mathcal{\tilde D}_{ij}$ are not in the top sectons. Thus, sectons and top pairs reveal different aspects of the correlations (see also Tab.~SIV). Finally, Fig.~\ref{fig:dcastat}(B) shows that sectons are also consistent with the decomposition into sectors, with almost no secton intersecting two sectors. 

Only few sectons are well-recognized structural or functional units: for the trypsin family, the top 6 sectons thus include 4 disulfide bonds~\footnote{Rat trypsin has a total of 6 disulfide bonds but the other 2 are not conserved in the family, with frequencies $<$ 5\%.}, and the catalytic triad, a group of three residues mediating peptide bond hydrolysis and shared among several other protein families~\cite{biochem}. Characterizing the structural and/or functional roles of other sectons is an open experimental challenge. Sectons are found in other protein families~\cite{RR}, thus raising the question of whether different families sharing a common fold also share common sectons~\cite{Mirny:1999hh}. Sectons and sectors are in any case distinct from previously recognized structural units such as secondary structures or 'foldons'~\cite{Panchenko:1996wr}, which consist of consecutive positions along the chain. 

Formally, sectors and sectons originate from exclusive parts of the spectrum of a common covariance matrix, $\bar C_{ij}^{ab}=\bar f_{ij}^{ab}-\bar f_i^a \bar f_j^b$, defined from the regularized frequencies, $\bar f_i^a=(1-\mu) f_i^a+\mu/(A+1)$ and
$\bar f_{ij}^{ab}=(1-\mu) f_{ij}^{ab}+\mu/(A+1)^2$, where $A=20$ is the number of amino acids. A parameter $\mu=1/2$ is introduced by DCA to define the coupling matrix $J=-\bar C^{-1}$ on which $\mathcal{D}_{ij}=\mathcal{D}_{ij}[J]$ relies~\cite{Morcos:2011jg}. This regularization is not required for SCA, but using $\mathcal{C}_{ij}=\mathcal{C}_{ij}[\bar C]$ with $\mu=1/2$ instead of $\mu=0$, which amounts to adding random sequences to the alignment, does not alter significantly sector identification (Fig.~S14). If $\bar C=\sum_k|k\rangle\lambda_k\langle k|$ denotes the spectral decomposition of $\bar C$ in the bra-ket notation, with ordered eigenvalues $\lambda_1\geq\dots\geq\lambda_L$, we can decompose $\bar C$ as  $\bar C=\bar C^++\bar C^-$, where
\begin{equation*}
\bar C^+=\sum_{k\leq k^*}|k\rangle\lambda_k\langle k|,\quad{\rm and}\quad \bar C^-=\sum_{k> k^*}|k\rangle\lambda_k\langle k|.
\end{equation*}
With for instance $k^*=100$, the sectors inferred by SCA from $\mathcal{C}_{ij}[\bar C^+]$ are indiscernible from those from $\mathcal{C}_{ij}[\bar C]$ (Fig.~S14). On the other hand, using  $\mathcal{D}_{ij}[J^-]$ with $J^-=-\sum_{k> k^*}|k\rangle\lambda_k^{-1}\langle k|$ instead of $\mathcal{D}_{ij}[J]$ in DCA~\footnote{$J^-$ is the inverse of $-C^-$ on its non-zero eigenspace, spanned by $\{|k\rangle\}_{k>k^*}$.}, not only do we recover the same contacts and sectons (Fig.~S15), but an additional  $\sim 20\%$ of them are found to be structurally connected (Fig.~S16). The association of sectors and sectons to different parts of the spectrum of $\bar C$ relates to random matrix theory, which indicates that both ends of the spectra of under-sampled empirical covariance matrices are statistically significant~\cite{Laloux:1999wr}. 
%A similar analysis of cross-correlations between price fluctuations of stocks in finance thus found the largest spectral modes to be supported by correlated stocks associated with business sectors, and the smallest modes to be localized on strongly (anti)correlated pairs of stocks~\cite{Plerou:2002im}. 
The spectral decomposition however does not account {\it per se} for the relation between sectors, contacts and sectons, and it may ultimately not be the most relevant decomposition of the correlations.

\begin{figure}[t]
\begin{center}
\includegraphics[width=\linewidth]{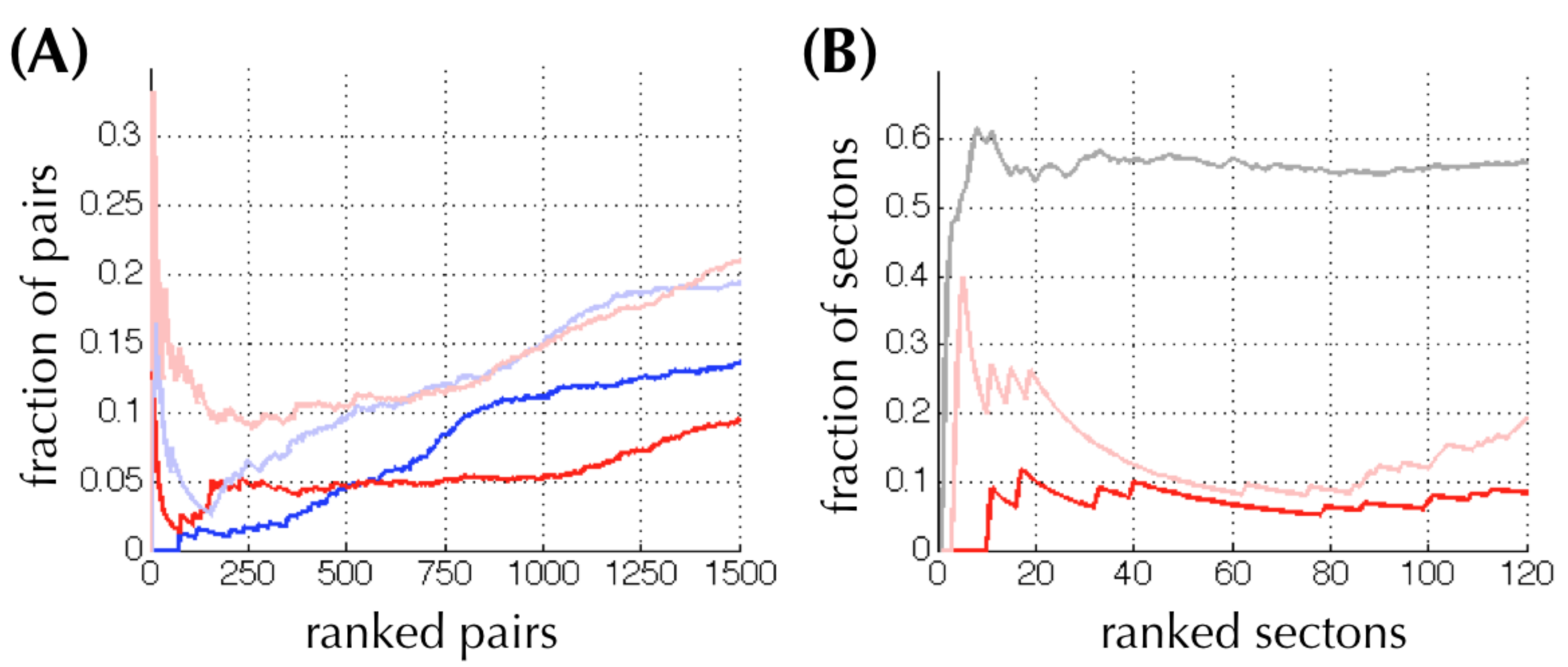}
\caption{Correlated pairs and sectons in bacterial genomes -- {\bf (A)}  Fraction of pairs of genes from different functional categories in the top pairs of $\mathcal{D}_{ij}$ (dark red curve) or $C_{ij}=f_{ij}-f_if_j$ (dark blue): up to rank 1000, a plateau around 5\% is observed in the first case, and a continuous increase in the second. The two matrices share some of their top pairs (Fig. S20), but, for instance, top pairs of $\mathcal{D}_{ij}$ initially contain more poorly characterized genes (light red) than top pairs of $C_{ij}$ (light blue). {\bf (B)} Fraction of sectons with two or more genes from different functional categories (dark red), and at least one poorly characterized gene (light red). In gray, fraction with genes from different functional categories after randomizing the content of the sectons, showing that finding less than 10\% of functionally mixed sectons is significantly lower.}
\label{fig:SectonsGenStat}
\end{center}
\end{figure}

The concepts and methods exposed thus far are not limited to protein structures. Another example involving biological sequences is the inference of functional couplings between genes in a genome. A first-order approach to this problem is to study the co-occurrence of genes in a large number genomes, also known as their phylogenetic profile~\cite{Pellegrini:1999ui}. The raw data is an $M\times L$ binary array $x_{si}$, where $x_{si}=1$ indicates that gene $i$ is present in the genome of species $s$, and 0 that it is absent ($A=1$ in this case). Building such a dataset requires mapping corresponding genes across genomes: here we rely on the partition of bacterial genes into clusters of orthologous genes (COGs)~\cite{Tatusov:2000tu}, to obtain a dataset consisting of $M\simeq 10^3$ genomes and $L\simeq 1.5$ $10^3$ orthologous classes~\cite{SI}.

No structural data is available for comparison in this case, but the classification of COGs into 3 broad, non- exclusive, functional classes [17] (metabolism, cellular processes, and information processing, with a 4th class for poorly characterized genes~\footnote{Gene annotation may however be incomplete and serve here only as a proxy for biological significance. Note also that COGs are themselves only proxies for orthologous classes, and more elaborated definitions of orthology may lead to more informative results.}) indicates that the top pairs of the matrix of direct information  $\mathcal{D}_{ij}$ are dominantly composed of genes from a same functional class (Fig.~\ref{fig:SectonsGenStat}(A)); these results are consistent with those previously derived from a similar approach
~\cite{Kim:2011}. As for protein alignments, sectons can be defined that consist here of small clusters, typically of 2 to 6 genes (Fig.~S17-20, Tab.~SV). These sectons are mostly composed of functionally related genes (Fig.~\ref{fig:SectonsGenStat}(B)); many sectons in fact consist of different subunits of a same protein complex (Tab.~SVI). Genomic sectors, involving larger groups of correlated genes, may be defined as well, although their significance is more difficult to assess (Fig.~S21, Tab.~SVII).

In conclusion, we provided evidence that the contacting pairs inferred by DCA~\cite{Morcos:2011jg} and the sectors inferred by SCA~\cite{Halabi:2009jca} are two interrelated features of a common pattern of coevolution, with coevolving units of intermediate size, called sectons, providing additional information. A fully unified mathematical framework for representing the hierarchy of correlations in biomolecules remains to be developed. Characterizing the structural, functional and evolutionary roles of patterns of coevolution is more generally a problem that extends beyond the scope of statistical studies of sequence data; in particular, experiments are needed to assess the extent to which statistical patterns of coevolution, inferred from a collection of sequences, are reflected in individual biomolecules.\\

Acknowledgements -- I thank L. Colwell, B. Houchmandzadeh, I. Junier, S. Kuehn, S. Leibler, R. Ranganathan, K. Reynolds, and T. Tesileanu for discussions and comments. This work is supported by ANR grant 'CoevolInterProt'.

\bibliographystyle{apsrev}

\onecolumngrid

\appendix

\newpage

\section{SUPPLEMENTAL MATERIAL}

\subsection{Preprocessing of the alignment}\label{sec:prepro}

As input for the identification of sectors and sectons in the trypsin family, we downloaded the full alignment PF00089 from Pfam (version 26.0, Nov. 2011). This alignment contains $M_0=14720$ sequences. It is represented by an array $x_{si}^a$ where $s$ labels the sequences (row in the alignment), $i$ the positions (columns) and $a$ is a number between 1 and 20 (each number is associated with one of the $A=20$ amino acids); $x_{si}^a=1$ indicates that sequence $i$ has amino acid $a$ at position $i$, and $x_{si}^a=0$ otherwise. As a reference for truncating the alignment and comparing to structural data, we used the sequence and structure of rat trypsin, chain E of the PDB id 3TGI in the Protein Data Base, which consists of $L_0=223$ positions.\\

To clean the alignment from an excess of gaps, the following operations were performed:

(1) Truncation of positions based on the reference sequence. As the alignment does not contain the last 7 positions of the reference sequence, this step leaves $L_1=216$ positions.

(2) Removal of sequences with a fraction of gaps exceeding $\gamma_{\rm seq}=0.2$, or with a sequence similarity to the reference sequence below $s_{\rm min}=0.2$, where sequence similarity is defined by 
\beq\label{eq:S}
S^{(1)}_{rs}=\frac{1}{L_1}\sum_{i=1}^{L_1}\sum_{a=1}^{20}x_{ri}^ax_{si}^a.
\eeq
This step leaves $M=9589$ sequences.

(3) Removal of positions with a fraction of gaps exceeding $\gamma_{\rm pos}=0.2$. The frequencies of gaps are computed with sequence weights as defined by Eq.~\eqref{eq:seqw}, using $S^{(1)}_{rs}$. This step leaves $L=204$ positions.\\

The parameters $\gamma_{\rm seq}=0.2$, $\gamma_{\rm pos}=0.2$ and $s_{\rm min}=0.2$ are chosen to mitigate the effects of gaps, but the results are not sensitive to their exact values. A more in-depth analysis of the structure of sequence correlations can reveal further information, and may suggest the removal of additional sequences, but this analysis is beyond the scope of the present study.

\subsection{Sequence-weighted frequencies}

Following Ref.~\cite{Weigt:2009we}, the uneven sampling of sequences is alleviated by introducing sequence weights defined by 
\beq\label{eq:seqw}
w_s\equiv \frac{\nu_s^{-1}}{\sum_r\nu_r^{-1}},\quad\textrm{with}\quad\nu_s\equiv |\{r: S_{rs}>\delta\}|,
\eeq
i.e., $\nu_r$ counts the number of sequences $r$ within distance $\delta$ of sequence $s$, where the distance between two sequences is the fraction of amino acids by which they differ,
\beq\label{eq:S}
S_{rs}=\frac{1}{L}\sum_{i=1}^{L}\sum_{a=0}^{20}\tilde x_{ri}^a \tilde x_{si}^a,
\eeq
where here, for numerical convenience, we include gaps by defining $\tilde x_{si}^a=x_{si}^a$ for $a>0$ and $x_{si}^0=1$ if sequence $s$ has a gap at position $i$.
 $M'=\sum_r\nu_r^{-1}$ can be interpreted as an effective number of sequences in the alignment. Here we take $\delta=0.8$, which results in $M'\simeq 4600$ effective sequences.\\

The sequence weights are used to define the frequency $f_i^a$ of amino acid $a$ at position $i$ and the joint frequency $f_{ij}^{ab}$ of amino acids $a,b$ at positions $i,j$ as
\beq\label{eq:f}
f_i^a\equiv\sum_sw_sx_{si}^a,\qquad f^{ab}_{ij}\equiv\sum_sw_sx_{si}^ax_{sj}^b.
\eeq
From these frequencies, a covariance matrix $C_{ij}^{ab}$ is defined by
\beq\label{eq:C}
C_{ij}^{ab}= f^{ab}_{ij}-f_i^af_j^b.
\eeq

\subsection{Regularization}

Direct coupling analysis (DCA) involves the inverse of a covariance matrix but $C_{ij}^{ab}$ is typically non invertible. A regularized covariance matrix is therefore introduced~\cite{Morcos:2011jg}, which is defined as
\beq\label{eq:regC}
\bar C_{ij}^{ab}=\bar f^{ab}_{ij}-\bar f_i^a\bar f_j^b,
\eeq
from the regularized frequencies
\beq
\bar f_i^a=(1-\mu) f_i^a+\mu\frac{1}{A+1},\qquad\bar f_{ij}^{ab}=(1-\mu) f_{ij}^{ab}+\mu\frac{1}{(A+1)^2},
\eeq
where $A=20$ for multiple sequence alignment (corresponding to the number of amino acids, with $A+1=21$ accounting for gaps), and $A=1$ for gene co-occurrence data. Here as in Ref.~\cite{Weigt:2009we}, $\mu=1/2$ is used, which effectively amounts to adding $M$ random sequences to the alignment.

\subsection{SCA matrix $\mathcal{C}_{ij}$}

Statistical Coupling Analysis (SCA) aims at identifying groups of positions under selection for a common functional property, based on two principles: the conservation of amino acids involved in the function, and their correlations induced by cooperative interactions~\cite{Lockless:1999uf}. SCA takes a heuristic approach to combine these two principles by weighting the covariance matrix $C_{ij}^{ab}$ of Eq.~\eqref{eq:C} with a measure of amino acid conservation $W_i^a$, defined by
\beq
W_i^a=\left|\ln\left(\frac{f_i^a(1-q^a)}{(1-f_i^a)q^a}\right)\right|,
\eeq
where $q^a=\sum_{i=1}^Lf_i^a/L$ is the mean frequency of amino acid $a$ (close in this case to the mean frequency inferred from a larger set of protein sequences used in Refs.~\cite{Lockless:1999uf,Halabi:2009jca}). Taking $W^i_aW^j_bC_{ij}^{ab}$ defines a conservation-weighted correlation matrix, which is reduced to a $L\times L$ correlation matrix between positions as follows~\cite{Lockless:1999uf,RR}:
\beq\label{eq:hatC}
\mathcal{C}_{ij}=\sqrt{\sum_{a,b=1}^A\left(W_i^aW_j^aC_{ij}^{ab}\right)^2}.
\eeq

For comparison with DCA, an equivalent quantity $\mathcal{\bar C}_{ij}$ is defined with same $W_i^a$ but the regularized covariance matrix $\bar C_{ij}^{ab}$ of Eq.~\eqref{eq:regC} instead of $C_{ij}^{ab}$.

\subsection{Matrix of direct information $\mathcal{D}_{ij}$}

In contrast to SCA, Direct Coupling Analysis (DCA) aims at identifying structural contacts between positions by inferring direct interactions from indirect correlations. It proceeds by a mapping to the problem of reconstructing the couplings $J_{ij}^{ab}$ of a $(A+1)$-state Potts model given its correlations $C_{ij}^{ab}$. Solving exactly this problem is computationally prohibitive, but mean-field approximations provide a range of alternatives. The simplest of these approximations consists in taking $J=-C^{-1}$. As the matrix $C$ is generally not invertible, the regularized covariance matrix $\bar C$ of Eq.~\eqref{eq:regC} is substituted for $C$. The couplings $J_{ij}^{ab}$ define for $i\neq j$ a model for the distribution of amino acids at every pair of positions $ij$,
\beq\label{eq:Pijc}
g_{ij}^{ab}=\exp (J_{ij}^{ab}+h_i^a+h_j^b+h_0),
\eeq
where $h_i^a$, $h_j^b$, $h_0$ are uniquely determined by requiring that $\sum_b g_{ij}^{ab}=\bar f^a_i$. From $g_{ij}^{ab}$, a matrix of direct information~\cite{Weigt:2009we} is defined by
\beq\label{eq:Dij}
\mathcal{D}_{ij}=\sum_{a,b=0}^A g_{ij}^{ab}\ln \frac{g_{ij}^{ab}}{\bar f_i^a \bar f^b_j}
\eeq
where the sum includes $a=0$ for gaps.

\subsection{Rotation by independent component analysis (ICA)}

Different implementations of ICA use different measures of independence and different algorithms for optimizing them. Here, we use one of the simplest implementations of ICA, called infomax~\cite{Bell:1995vn} (with modifications introduced in Ref.~\cite{Amari96}). We take as input the top $k$ eigenvectors of the correlation matrix $\mathcal{C}_{ij}$ or $\mathcal{\tilde D}_{ij}$, which we concatenate in a $k_{\rm top}\times L$ matrix $Z$ (at variance with usual implementations of ICA which take the dataset $X$ as input). The algorithm iteratively updates an unmixing matrix $W$, starting from the $k_{\rm top}\times k_{\rm top}$ identity matrix $W_0=I_{k_{\rm top}}$, with increments $\Delta W$ given by
\beq
\Delta W=\eta\left(I_{k_{\rm top}}+\left(1-\frac{2}{1+\exp (-WZ)}\right)(WZ)^\top\right)W.
\eeq
The parameter $\eta$ is a learning rate that has to be sufficiently small for the iterations to converge. In this study, $\eta=10^{-4}$ led to convergence after $10^4$ iterations in applications to the trypsin alignment, and $\eta=10^{-5}$ after $10^5$ iterations in applications to the co-occurrence of genes in bacterial genomes. \\

The independent components $V^{(k)}$ are obtained by applying $W$ to the eigenvectors in $Z$. To set their overall scale and sign, we normalize them to unit length ($\sum_i (V_i^{(k)})^2=1$) and orient them so that the position $i$ with largest $|V^{(k)}_i|$ satisfies $V^{(k)}_i>0$. The order of the independent components, which is not necessarily prescribed in other implementations of ICA, is here well defined by the algorithm and is related to the order of the principal components. 

\subsection{Threshold $k_{\rm top}$ in defining sectors}

The spectrum of $\mathcal{C}_{ij}$, displayed in Fig.~S\ref{fig:spectrum}, indicates that between 3 to 7 eigenvalues are emerging from a bulk of small eigenvalues. This estimation is confirmed by comparing with the spectra of randomized alignments, where the amino acids are drawn independently at each position $i$ according the frequencies $f_i^a$, so as to remove the correlations but preserve the distribution of amino acids at each position.\\

In the main text, we presented the results when selecting $k_{\rm top}=4$ modes. A smaller number of components may prevent the discrimination between sectors, as shown in Fig.~S\ref{fig:sect3}, where taking $k_{\rm top}=3$ causes the red and blue sectors to appear along a same component. Reciprocally, a larger number of components may lead to the splitting of a sector into disconnected subsets, as shown in Fig.~S\ref{fig:sect5}, where taking $k_{\rm top}=5$ decomposes the purple sector along two components. In this case, the two components do not define two new sectors, but indicate a partition of the sector into a core and a periphery, as shown in Fig.~S\ref{fig:sect5more}. 

\subsection{Threshold $\epsilon$ in defining sectors}

Besides the threshold $k_{\rm top}$ for the eigenvalues, the definition of sectors involves a threshold $\epsilon$ determining which positions contribute significantly to each component. Here again, $\epsilon$ can be estimated from a comparison with randomized alignments, but it is more interesting to notice that several values of $\epsilon$ are consistent with structurally connected sectors. Varying $\epsilon$ thus defines a hierarchy of structurally connected positions, from the core of the most conserved positions for large $\epsilon$ to a periphery of less conserved positions for smaller $\epsilon$.\\

As an illustration of this feature, we show in Fig.~S\ref{fig:epsilon} how the connectivity of each sector, measured by the relative size of its largest structurally connected subset, varies with $\epsilon$. With the possible exception of the cyan sector, the sectors are found to be significantly structurally connected for nearly all values of $\epsilon$. The significance of this finding is assessed by comparing with randomly-formed groups of positions, or with the positions ordered by their overall degree of conservation (Fig.~S\ref{fig:epsilon}(C)).

\subsection{Composition of sectors}

Tab.~S\ref{fig:sectcompo} reports the exact composition of the 4 sectors defined in the main text. The green and red sectors have very significant overlap with the green and red sectors previously defined in Ref.~\cite{Halabi:2009jca}. On the other hand, the purple and cyan sectors have only limited overlap with the blue sector defined in this previous study. As a visual representation of the relation between these two definitions, Fig.~S\ref{fig:sectors_oldnew} reproduces Fig.~1, but with colors corresponding to the 3 sectors defined in Ref.~\cite{Halabi:2009jca}. While the implementation of SCA is slightly different in the two studies, the difference mainly stems from the difference between the alignments that serve as inputs: the present study is based on an alignment with nearly 10 more sequences (and $L=204$ positions instead of 223). \\

In Fig.~1, we eluded the discussion of positions at the intersection between different $\S_k=\{i : V^{(k)}_i>\epsilon\}$ by  excluding them from the definition of sectors. These few positions are however structurally meaningful: Fig.~S\ref{fig:sectinter} shows that they are at the periphery of one of the two sectors, and, in several instances, at the structural interface between them.

\subsection{Contacts and sectons}

Fig.~S\ref{fig:truepos} reports the performance of DCA for predicting contacts for the alignment under study: up to rank 79, all top pairs of positions in terms of $\mathcal{\tilde D}_{ij}$ are actual contacts. Tab.~S\ref{fig:sectoncontact} gives the list of the top 120 pairs and indicates the secton to which they belong, when applicable. Tab.~S\ref{fig:sectons_list} gives the composition of the top 120 sectons and indicates which are connected. Fig.~S\ref{fig:sectonStruct} shows the top 36 sectons on the three-dimensional structure of trypsin: the 36th secton is the first to include a position disconnected from the others.\\

Fig.~S\ref{fig:secton_connect} shows that the top 35 sectons are structurally connected (and slightly more if considering a more stringent cutoff $\epsilon$). It also shows that, as for sectors, the results are insensitive to the choice of the threshold $\epsilon$ used in defining the positions contributing significantly to a component. In all figures involving sectons, $k_{\rm top}=120$, except Fig.~S\ref{fig:firstsecton}, which consider varying $k_{\rm top}$:  the definition of sectons is found to be insensitive to the value of $k_{\rm top}$, except for the first secton, which appears to be peculiar and unstable. For instance, the first secton is the only secton of size 2 that is not in the top pairs of $\mathcal{\tilde D}_{ij}$, and is responsible for the blue curve not reaching 1 in Fig.~2(C).\\

Not truncating $\mathcal{D}_{ij}$ to $\mathcal{\tilde D}_{ij}$ leads to the sectons shown in Fig.~S\ref{fig:DInotr}. Many of these sectons consist of consecutive positions. These trivial sectons are induced by gaps, which tend to be consecutive along the sequence (this feature is partly a consequence of the multiple sequence alignment algorithm, which have a penalty for opening new gaps). Truncating $\mathcal{D}_{ij}$ (or equivalently $J_{ij}^{ab}$) is a simple if not optimal way of getting rid of these trivial correlations.\\

Figs. 2(C-D) show that the top pairs of $\mathcal{\tilde D}_{ij}$ and sectons contain overlapping but non-equivalent informations. This is also shown by Tab.~\ref{fig:subgraphs}, which analyzes the connected subgraphs of the contact graph inferred from the top pairs of $\mathcal{\tilde D}_{ij}$: these connected subgraphs are related, but non-identical to the sectons.

\subsection{Orthogonal decomposition of the correlations}

As in the main text, $\bar C=\sum_k|k\rangle\lambda_k\langle k|$ denotes the spectral decomposition of the regularized covariance $\bar C$  in the bra-ket notation, with ordered eigenvalues $\lambda_1\geq\dots\lambda_L$. This decomposition is used to defined the two matrices
\beq
\bar C^+=\sum_{k\leq k^*}|k\rangle\lambda_k\langle k|,\quad{\rm and}\quad \bar C^-=\sum_{k> k^*}|k\rangle\lambda_k\langle k|.
\eeq

Fig.~S\ref{fig:sectorstop} shows the differences between the components of the SCA matrices defined from $C$ (the original covariance matrix with $\mu=0$ and $k^*=0$), $\bar C$ (the regularized covariance matrix with $\mu=1/2$ and $k^*=0$) and $\bar C^+$ (with $\mu=1/2$ and $k^*=100$). Keeping only the top $k^*$ modes has no incidence on the top components of the SCA matrix used to define sectors, while regularizing has only a minor effect.\\

The sectons defined from $J^-=-\sum_{k> k^*}|k\rangle\lambda_k^{-1}\langle k|$, with again $k^*=100$, are comparable to those defined when including all the modes, as shown in Fig.~S\ref{fig:sectonsbot} (with the first secton forming an exception, as in Fig.~S\ref{fig:firstsecton}). Fig.~S\ref{fig:lessismore} shows that truncating the top $k^*$ modes of $\bar C$ can actually cause more top pairs and more sectons to be structurally connected.

\subsection{Co-occurrence of orthologous genes in bacterial genomes}

Sequenced bacterial genomes and COG annotations were downloaded from NCBI. The initial dataset contained $M_0=1432$ genomes and $L_0=4467$ COGs.\\

The following cleaning steps were conducted:

(1) Removal of 'exceptional' genomes with size below 500 kpb or with no less than 60 \% of genes annotated by COGs. This step leaves $M=1108$ genomes.

(2) Removal of COGs that are present in less than $\gamma_c=0.4$ of the genomes, where gene frequencies are computed with sequence weights using $\delta=0.9$. The relatively high value $\gamma_c=0.4$ is meant to reduce the data to a size that is easily tractable computationally; here it leaves $L=1474$ COGs. Conversely, the choice of $\delta=0.9$ is meant to preserve a relatively high effective number of genomes, here $M'=380$. The exact values of these parameters are however not crucial.\\

The data is represented by a $M\times L$ binary array $x_{si}$ with $x_{si}=1$ if genome $s$ has at least one gene in COG $i$, and 0 otherwise. The average occurrence of genes is $q=\sum_{si}x_{si}/(ML)=0.67$. This dataset is in no way meant to be optimal, and finer definitions of orthology are possible. Our point here is to show that sectors and sectons can be unraveled even from a relatively crude construction of bacterial phylogenetic profiles, leaving for future work the study of more elaborated datasets.\\

Genomic sectons are obtained exactly as protein sectons, except that the alphabet is now binary ($A=1$), sequence weights are computed with $\delta=0.9$, and $\mathcal{D}_{ij}$ is not truncated ($\Delta=0$, with $\mathcal{D}_{ii}=0$). The covariance matrix $C_{ij}=f_{ij}-f_if_j$ is computed from the frequencies $f_i$ and $f_{ij}$ of occurrence of genes and co-occurrence of pairs of genes. The content of first 100 sectons (obtained with $k_{\rm top}=120$) is reported in Tab.~S\ref{fig:gensectons_list}, with further details for the top 24 sectons provided in Tab.~S\ref{tab:sectons2}.\\ 

Fig.~S\ref{fig:sizeSectGen} reports the distribution of sizes of genomic sectons and Fig.~S\ref{fig:ComparCDS} analyzes the differences of content between top pairs of $\mathcal{D}_{ij}$, top pairs of $C_{ij}$, and sectons. Finally, Fig.~S\ref{fig:SectonC} shows that applying ICA to the top $k_{\rm top}=120$ eigenvectors of $C_{ij}$ does not lead to localized components as it does when applying it to $\mathcal{D}_{ij}$ (Fig.~S\ref{fig:gensectons}).\\

Genomic sectors can also be defined following the methods for defining protein sectors. Fig.~S\ref{fig:SectorsGen} shows the counterpart of Fig.~1, using here $k_{\rm top}=6$. In absence of a counterpart for the experimentally determined three-dimensional structure of proteins, assessing the relevance of these sectors is not obvious. Using the partition of COGs into 3 broad functional classes~\cite{Tatusov:2000tu}, we nevertheless find that 4 of the 6 components support groups of COGs that are significantly enriched in some of these classes, as reported in Tab.~S\ref{fig:SectorsGenStat}. This suggests that genome sectors may be defined as well, which contain co-functional and therefore coevolving genes.

\clearpage

\setcounter{figure}{0}

\begin{figure}[t]
\renewcommand{\figurename}{FIG. S}\begin{center}
\includegraphics[width=.8\linewidth]{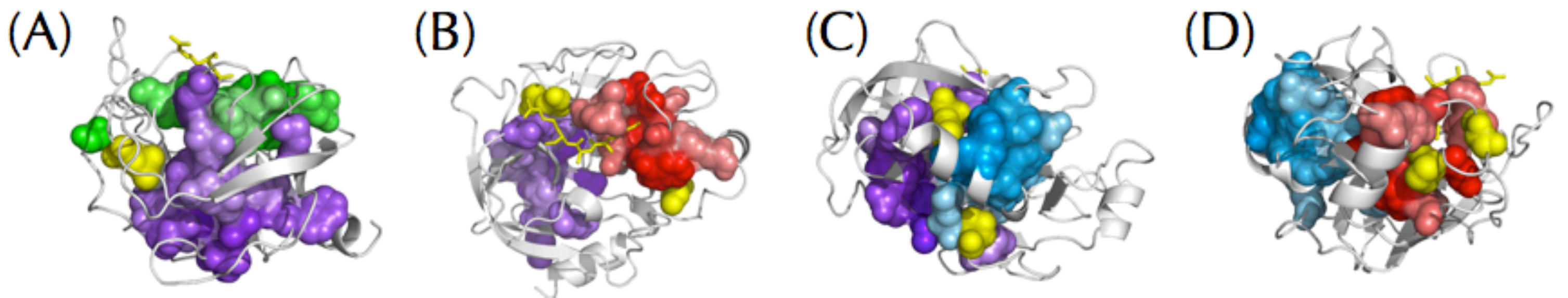}
\caption{Intersections between $\S_k$ -- In the main text, sectors are defined by $\S'_k=\{i:V_i^{(k)}>\epsilon,  V_i^{(\ell)}<\epsilon,\ \ell\neq k\}$ ($k=1,\dots,k_{\rm top}=4$, $\epsilon=0.1$). Here, we represent in yellow the few positions that belong to the intersections $\S_k\cap\S_\ell$ for $k\neq\ell$, with $\S_k=\{i:V_i^{(k)}>\epsilon\}$. There are 7 such positions, all of which are at the structural periphery of either $\S'_k$ or $\S'_\ell$, including 4 that are at the structural interface between the two. (A position from the green sector appears as disconnected in (A), but the position that could connect it to the rest of the green sector is actually not included in the alignment).}
\label{fig:sectinter}
\end{center}
\end{figure}

\begin{figure}[t]
\renewcommand{\figurename}{FIG. S}
\begin{center}
\includegraphics[width=.7\linewidth]{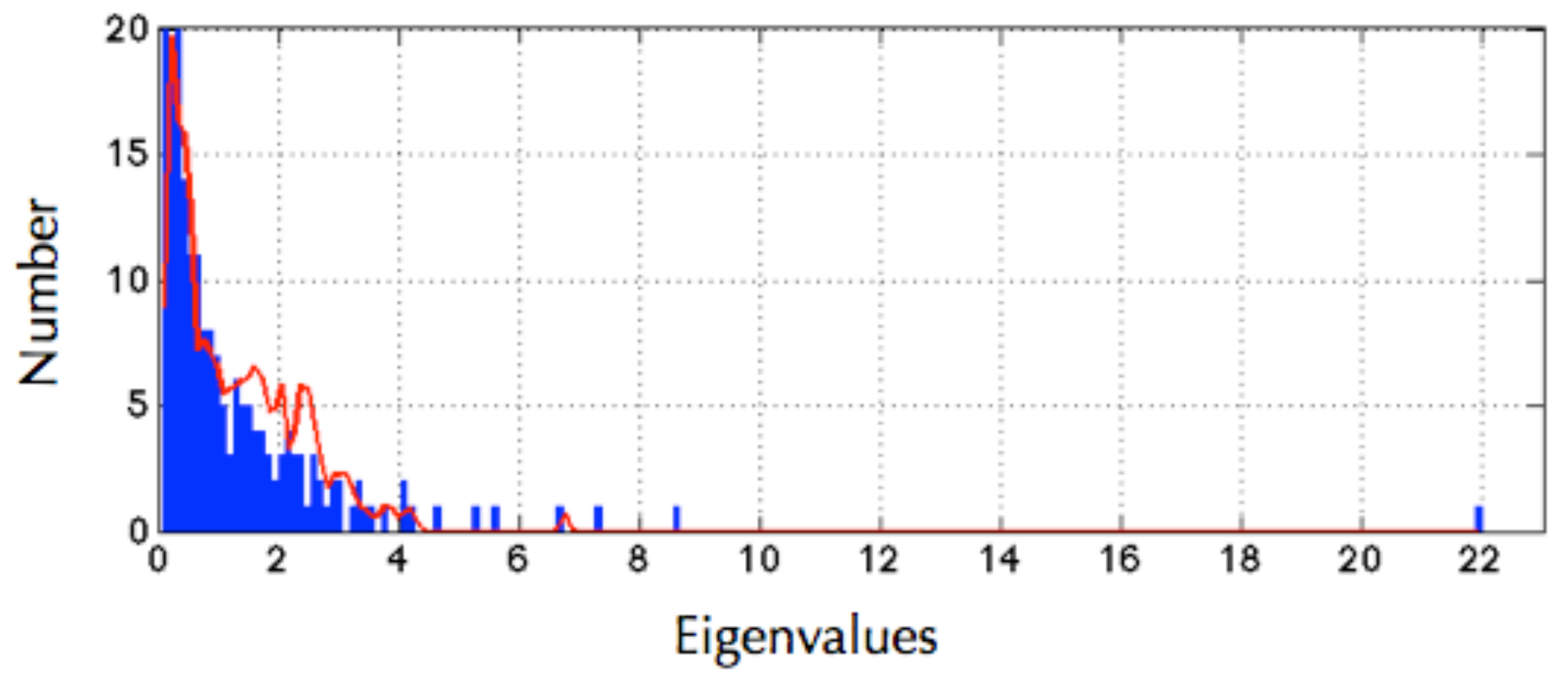}
\caption{Spectrum of the SCA matrix $\mathcal{C}_{ij}$ -- In blue, histogram of the $L$ eigenvalues of the matrix $\mathcal{C}_{ij}$ (truncated to 20 along the $y$-axis). In red, average spectrum over 100 randomized alignments, where the amino acids are drawn independently at each position $i$ according to the frequencies $f_i^a$. This shows that between 3 and 7 eigenvalues may considered as significant.}
\label{fig:spectrum}
\end{center}
\end{figure}

\begin{figure}[t]
\renewcommand{\figurename}{FIG. S}\begin{center}
\includegraphics[width=.65\linewidth]{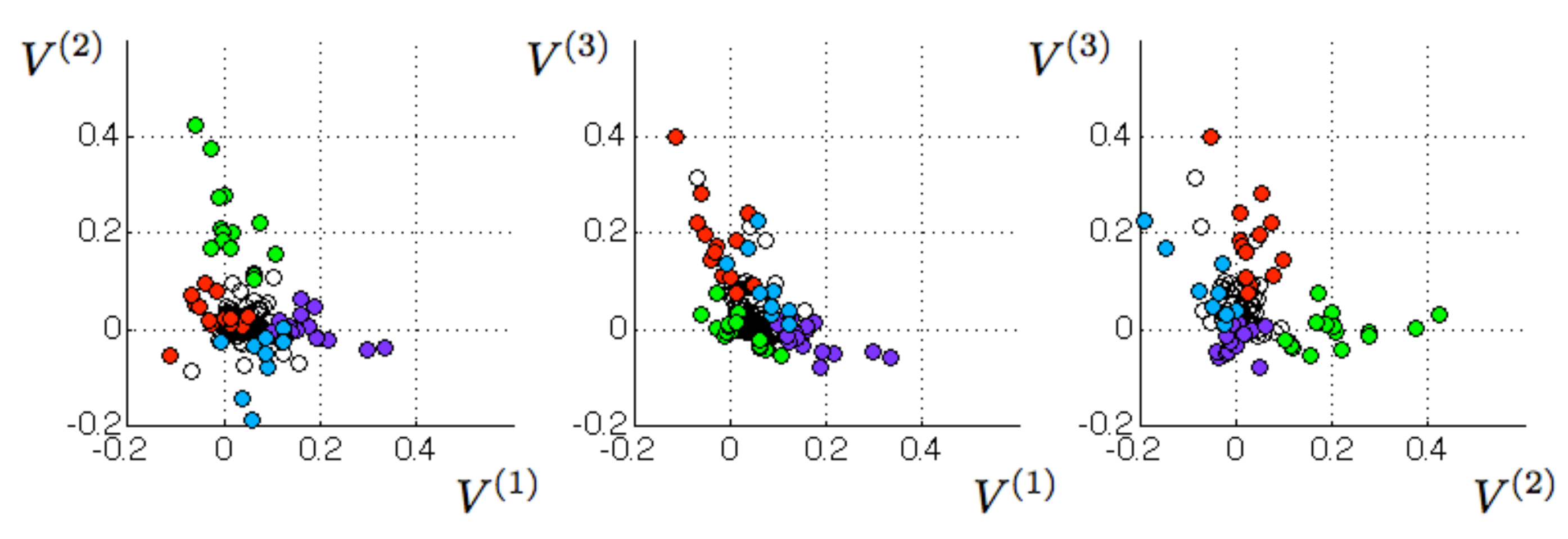}
\caption{Independent components from rotation by ICA of the top $k_{\rm top}=3$ eigenvectors of $\mathcal{C}_{ij}$ -- This figure is the counterpart of Fig.~1, for $k_{\rm top}=3$ instead of $k_{\rm top}=4$, and with positions colored as in Fig.~1. The same green and red sectors are defined along $V^{(1)}$ and $V^{(2)}$, but the red and cyan sectors appear together along $V^{(3)}$.}
\label{fig:sect3}
\end{center}
\end{figure}

\begin{figure}[t]
\renewcommand{\figurename}{FIG. S}\begin{center}
\includegraphics[width=.65\linewidth]{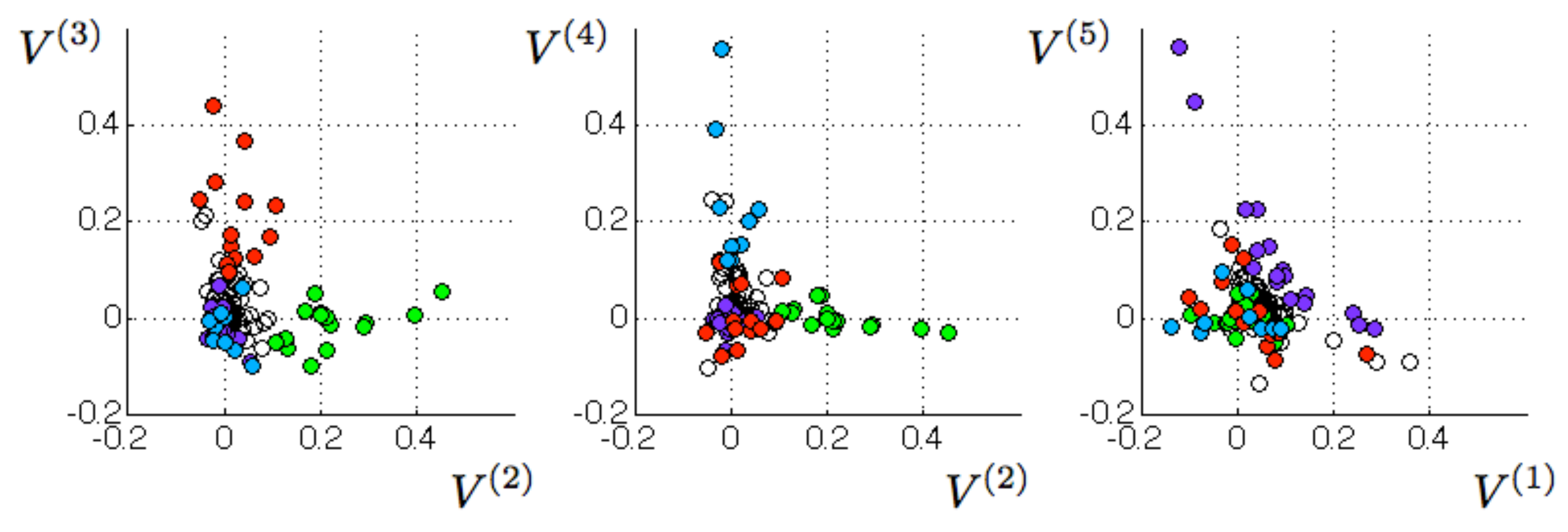}
\caption{Independent components from rotation by ICA of the top $k_{\rm top}=5$ eigenvectors of $\mathcal{C}_{ij}$ -- This figure is the counterpart of Fig.~1, for $k_{\rm top}=5$ instead of $k_{\rm top}=4$, and with positions colored as in Fig.~1. The definitions of the green, red and cyan sectors along $V^{(2)}, V^{(3)}, V^{(4)}$ are consistent with their definition with $k_{\rm top}=4$, while the new component $V^{(5)}$ decomposes the purple sector in two subsets, whose interpretation is given in Fig.~S\ref{fig:sect5more}.}
\label{fig:sect5}
\end{center}
\end{figure}

\begin{figure}[t]
\renewcommand{\figurename}{FIG. S}\begin{center}
\includegraphics[width=.65\linewidth]{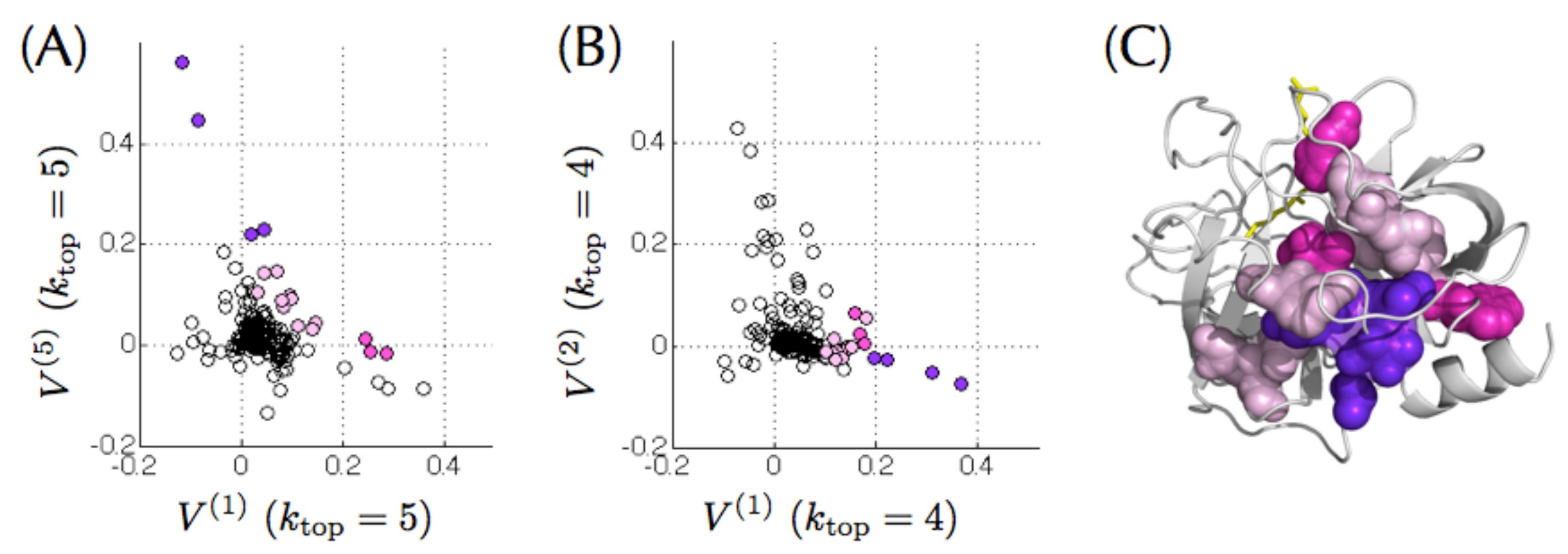}
\caption{Interpretation of the decomposition of the purple sector when considering $k_{\rm top}=5$ -- {\bf (A)} Same as the last graph of Fig.~S\ref{fig:sect5}, except that the positions from the purple sector are indicated with three different colors, depending on whether they satisfy $V^{(1)}_i>0.2$ (pink), $V^{(5)}_i>0.2$ (purple) or neither (light pink). {\bf (B)} Using the same coloring scheme, but now for the projection along components obtained with $k_{\rm top}=4$ as in Fig.~1, shows that the partition of the purple sector corresponds to a partition between a core, defined as the positions with highest contribution to $V^{(1)}$, and other positions. {\bf (C)} Location of the positions on the three-dimensional structure, showing that the core is structurally connected with the other positions around.} %Peeling the onion.}
\label{fig:sect5more}
\end{center}
\end{figure}

\begin{figure}[t]
\renewcommand{\figurename}{FIG. S}\begin{center}
\includegraphics[width=.65\linewidth]{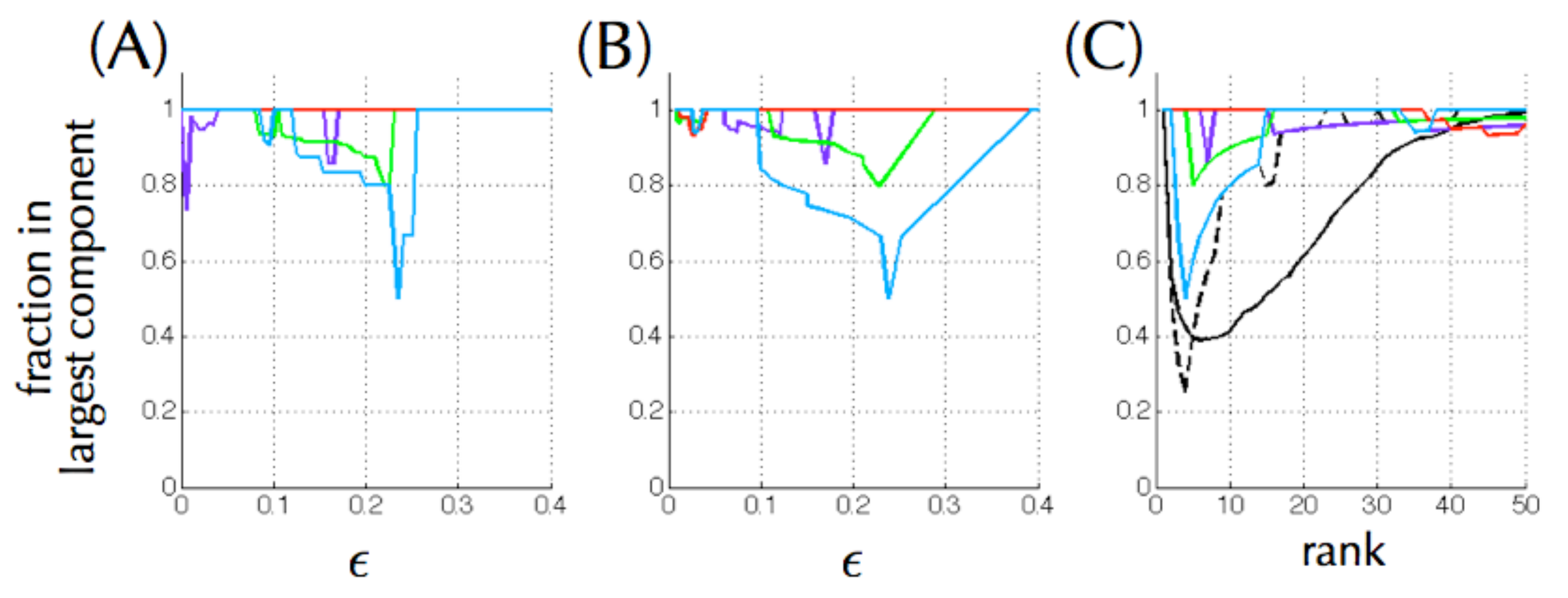}
\caption{Connectivity of the sectors for varying values of $\epsilon$ -- {\bf (A)} Fraction of the positions in the largest structurally connected subset of a sector for varying values of $\epsilon$ and for sectors defined as in the main text by $\S'_k=\{i:V_i^{(k)}>\epsilon,  V_i^{(\ell)}<\epsilon,\ \ell\neq k\}$ ($k=1,\dots,k_{\rm top}=4$). The colors correspond to those of Fig.~1: magenta for $k=1$, green for $k=2$, red for $k=3$ and cyan for $k=4$. {\bf (B)} Same as (A) but not excluding intersections, which are significant for small $\epsilon$, i.e., taking $\S_k=\{i:V_i^{(k)}>\epsilon\}$. {\bf (C)} To assess the significance of (A) and (B), we consider also two other groups of ordered positions: randomly ranked positions and positions ranked by degree of conservation, measured by the relative entropy $D_i=\sum_{a=0}^{20} f_i^a\ln f_i^a/q^a$ with $f_i^0=1-\sum_{a=1}^{20} f_i^a$ for the frequency of gaps at position $i$ and $q^a=\sum_i f_i^a/L$ for the background frequencies. The results are presented here as a function of the size of the groups, where positions are added according to their rank (given by $V_i^{(k)}$ for the sectors). Compared to randomly ranked positions (full black line), sectors are clearly significantly more connected. They are also more connected than positions ranked by conservation (dashed black line), with the possible exception of the cyan sector.}
\label{fig:epsilon}
\end{center}
\end{figure}

\begin{table}[t]
\renewcommand{\tablename}{TAB. S}\begin{tabular}{|c|c|c|c|}
  \hline
\ Purple sector\ &\ Green sector\ &\ Red sector\ &\ Cyan sectorÊ \\
  \hline
29 & 57$^*$ & 228$^*$ & 237$^*$ \\
122 & 195$^*$ & 189$^*$ & 238 \\
46$^*$ & 197$^*$ & 215$^*$ & 234 \\
120 & 55$^*$ & 183$^*$ & 91 \\
28 & 19$^*$ & 226$^*$ & 231 \\
51 & 102$^*$ & 216$^*$ & 103 \\
40 & 43$^*$ & 213$^*$ & 229$^*$ \\
30 & 194$^*$ & 227$^*$ & 101 \\
104$^*$ & 193 & 184 & 123$^*$ \\
27 & 214$^*$ & 191$^*$ & 199 \\
31 & 94 & 192$^*$ &  \\
136$^*$ & 142$^*$ & 172$^*$ &  \\
52$^*$ & 42$^*$ & 138 &  \\
201$^*$ & 58$^*$ &  &  \\
32 & 33$^*$ &  &  \\
68$^*$ &  &  &  \\
200 &  &  &  \\
\hline
\end{tabular}
\caption{Composition of the sectors defined in Fig.~1, ranked by the corresponding value of $V^{(k)}_i$ -- A star indicates for the purple and cyan sectors that the position belongs to the blue sector defined in Ref.~\cite{Halabi:2009jca}, and for the green and red sectors that it belongs respectively to the green and red sector of Ref.~\cite{Halabi:2009jca}. Note that 19 positions included in the alignment of Ref.~\cite{Halabi:2009jca} are not represented in the present alignment, including 1 of the 20 positions of the red sector defined in Ref.~\cite{Halabi:2009jca}, 0 of the 22 positions of its green sector and 4 of the 25 positions of its blue sector.}\label{fig:sectcompo}
\end{table}

\begin{figure}[t]
\renewcommand{\figurename}{FIG. S}\begin{center}
\includegraphics[width=.6\linewidth]{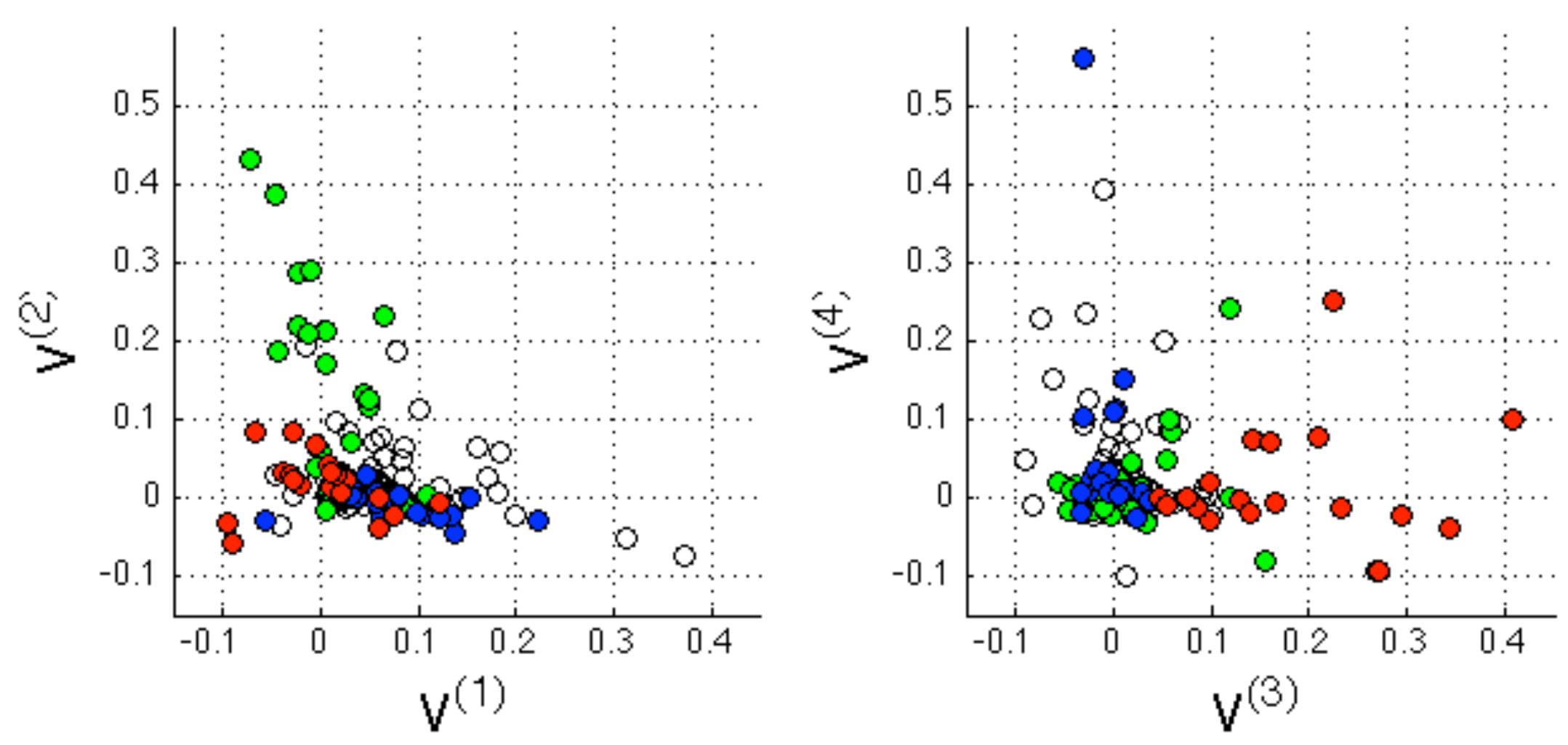}
\caption{Comparison with the sectors defined in Ref.~\cite{Halabi:2009jca} -- The graphs are identical to those of Fig.~1(A), except that the positions are colored according to the definition of the 3 sectors of Ref.~\cite{Halabi:2009jca}. This shows as in Tab.~S\ref{fig:sectcompo}
that essentially the same green and red sectors are identified, while the purple and cyan sectors have a small overlap with the previously defined blue sector.}
\label{fig:sectors_oldnew}
\end{center}
\end{figure}

\clearpage

\begin{table}[t]
\renewcommand{\tablename}{TAB. S}
 \begin{minipage}[r]{.33\linewidth}
{\small
\begin{tabular}{|c|c|c|c|}
  \hline
contact & pair & secton(s) & dist. (\AA) \\
  \hline
1&42 - 58 & 2 & 2 \\
2&136 - 201 & 5 & 2 \\
3&57 - 195 & 3 & 2.7 \\
4&191 - 220 & 6 & 2.2 \\
5&168 - 182 & 4 & 2 \\
6&32 - 40 & 7 & 2.8 \\
9&189 - 226 & 8 & 3.3 \\
11&59 - 104 & 11 & 3.9 \\
16&29 - 122 & 15 & 6.7 \\
17&26 - 157 & 10 & 4.9 \\
20&44 - 52 & 9 & 4.3 \\
21&142 - 194 & 12 & 2.7 \\
25&30 - 139 & 13 & 2.7 \\
36&51 - 105 & 16 & 3.8 \\
37&46 - 112 & 14 & 4 \\
38&27 - 157 & 10 & 3.8 \\
40&161 - 184 & 19 & 3.1 \\
44&146 - 221A & 21 & 3.6 \\
54&68 - 112 & 14 & 3.8 \\
58&51 - 107 & 16 & 3.6 \\
59&53 - 209 & 17 & 3.5 \\
64&57 - 102 & 3 & 2.7 \\
71&21 - 156 & 24 & 3.3 \\
75&31 - 44 & 9 & 3.1 \\
76&86 - 107 & 22 & 2.7 \\
77&45 - 209 & 17 & 3.8 \\
78&100 - 179 & 25 & 3 \\
80&158 - 188A & 26 & 3.5 \\
83&138 - 190 & 18 & 4.7 \\
84&190 - 213 & 18 & 3.7 \\
99&189 - 228 & 44 & 3.9 \\
104&138 - 213 & 18 & 4.2 \\
106&102 - 195 & 3 & 6.1 \\
112&103 - 234 & 20 & 4.1 \\
114&22 - 157 & - & 2 \\
119&31 - 68 & 84 & 3.8 \\
123&31 - 52 & 9 & 4 \\
125&30 - 155 & - & 4 \\
126&180 - 215 & 23 & 3.8 \\
127&215 - 227 & 23 & 2.8 \\
 \hline
\end{tabular}
}
\end{minipage} \hfill
\begin{minipage}[c]{.32\linewidth}
{\small
\begin{tabular}{|c|c|c|c|}
\hline
contact & pair & secton(s) & dist. (\AA) \\
\hline
128&27 - 139 & - & 3.6 \\
129&189 - 221 & 44 & 3 \\
130&81 - 118 & 28 & 3.8 \\
134&136 - 162 & - & 3.3 \\
135&136 - 199 & - & 3.8 \\
137&165 - 230 & 32 & 5 \\
141&50 - 107 & 22 & 3.3 \\
142&183 - 189 & - & 4.7 \\
144&29 - 200 & - & 3.5 \\
145&46 - 68 & 14 & 3.9 \\
146&183 - 199 & 27 & 4.3 \\
147&138 - 158 & 26 & 2.8 \\
148&136 - 210 & - & 5.4 \\
153&45 - 121 & 17 31 & 4.1 \\
154&189 - 216 & - & 6.2 \\
155&121 - 209 & 17 & 4 \\
156&195 - 214 & - & 3.4 \\
159&46 - 52 & - & 3.8 \\
161&100 - 177 & 59 & 2.6 \\
164&67 - 82 & 35 & 3.5 \\
167&103 - 229 & 20 & 4.2 \\
169&181 - 199 & 27 & 6.6 \\
171&139 - 200 & - & 6.7 \\
173&114 - 120 & 33 & 3.9 \\
174&55 - 195 & - & 4.7 \\
175&85 - 106 & 29 & 3.4 \\
178&91 - 101 & 95 & 3.6 \\
182&101 - 234 & 95 & 2.6 \\
184&84 - 109 & 38 & 2.9 \\
189&50 - 111 & 37 & 3.6 \\
192&102 - 214 & 41 & 2.5 \\
199&195 - 213 & - & 3.7 \\
200&142 - 193 & - & 3.7 \\
201&27 - 137 & 39 & 4.4 \\
203&87 - 107 & 85 & 2.8 \\
209&46 - 108 & - & 4.3 \\
211&195 - 216 & - & 7.8 \\
214&161 - 184A & - & 3.8 \\
217&179 - 233 & 25 & 3 \\
218&182 - 191 & - & {\color{red} 13} \\
\hline
\end{tabular}
}
\end{minipage} \hfill
\begin{minipage}[l]{.32\linewidth}
{\small
\begin{tabular}{|c|c|c|c|}
\hline
contact & pair & secton(s) & dist. (\AA) \\
\hline
223&138 - 160 & 98 & 4.1 \\
227&100 - 180 & - & 3.8 \\
231&46 - 114 & - & 3.6 \\
233&124 - 235 & 34 & 3.8 \\
235&72 - 78 & 40 & 8 \\
236&30 - 198 & 13 & 3.3 \\
240&183 - 226 & - & 3.2 \\
241&27 - 122 & - & {\color{red} 11} \\
243&83 - 112 & - & 4 \\
244&57 - 213 & - & 4.5 \\
245&139 - 157 & - & 4.2 \\
250&190 - 216 & 45 & 7.1 \\
251&68 - 118 & - & 4 \\
256&57 - 193 & - & 7.7 \\
261&144 - 152 & 43 & 3.8 \\
264&140 - 194 & - & 3.3 \\
266&45 - 212 & 30 & 6.3 \\
268&101 - 179 & - & 3.3 \\
270&216 - 227 & 45 & 4.7 \\
273&49 - 111 & 37 & 3.2 \\
274&163 - 225 & 42 & 3.7 \\
280&43 - 57 & - & {\color{red} 8.3} \\
285&139 - 198 & 13 & 3.6 \\
286&55 - 197 & 36 & 5.7 \\
289&219 - 226 & - & 6.6 \\
292&21 - 154 & 89 & 3.8 \\
293&86 - 109 & - & 3.9 \\
296&30 - 43 & - & 4 \\
298&179 - 234 & - & 3.2 \\
300&122 - 189 & - & {\color{red} 21} \\
301&43 - 142 & - & 7.8 \\
309&83 - 109 & - & 3.8 \\
311&90 - 104 & - & 3.9 \\
314&199 - 228 & - & 3.3 \\
321&160 - 228 & 98 & 6.8 \\
328&29 - 124 & - & {\color{red} 12} \\
333&180 - 189 & - & {\color{red} 13} \\
334&135 - 161 & - & 3.2 \\
337&121 - 200 & 31 118 & 3.9 \\
338&180 - 227 & 23 & 3.8 \\
\hline
\end{tabular}
}
\end{minipage} 
\caption{Top 120 non-trivial contacts and their associated secton -- The pairs are ordered by decreasing values of $\mathcal{D}_{ij}$, with their rank indicated in the first column. Only pairs of positions separated by $>\Delta=5$ amino acids along the sequence are considered, and many pairs are therefore not included (red curve in Fig.~S\ref{fig:truepos}). The second column indicates the rank of the secton where the pair is found and the third the distance in angstroms between the positions in the three-dimensional structure. Pairs are considered in contact when this distance is $<8$ \AA, and false positive are indicated in red. Here, none of these false positive are in a secton, but many true positive are also not in a secton.}\label{fig:sectoncontact}        
\end{table}

\clearpage

\begin{figure}[t]
\renewcommand{\figurename}{FIG. S}\begin{center}
\includegraphics[width=.8\linewidth]{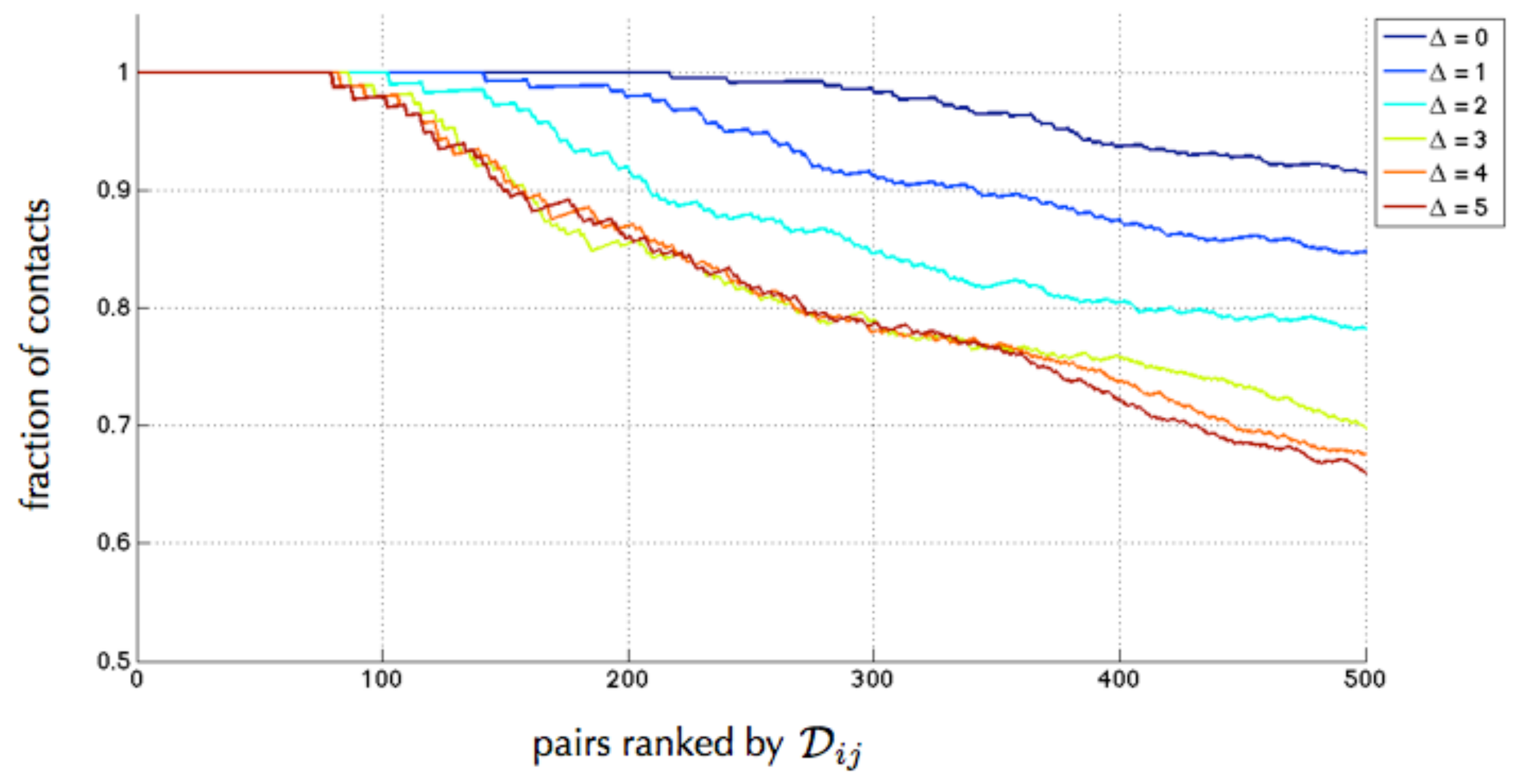}
\caption{Fraction of pairs of positions predicted to be in contact from the top values of the matrix of direct information $\mathcal{D}_{ij}$ -- Pairs $ij$ of positions are ordered by decreasing values of $\mathcal{D}_{ij}$, and the fraction of top $n$ pairs to be in structural contact (distance $<8$ \AA) is considered as a function of $n$. Different curves correspond to different values of $\Delta$, where pairs separated by $\leq\Delta$ amino acids along the sequence are ignored.}
\label{fig:truepos}
\end{center}
\end{figure}

\begin{figure}[t]
\renewcommand{\figurename}{FIG. S}\begin{center}
\includegraphics[width=.6\linewidth]{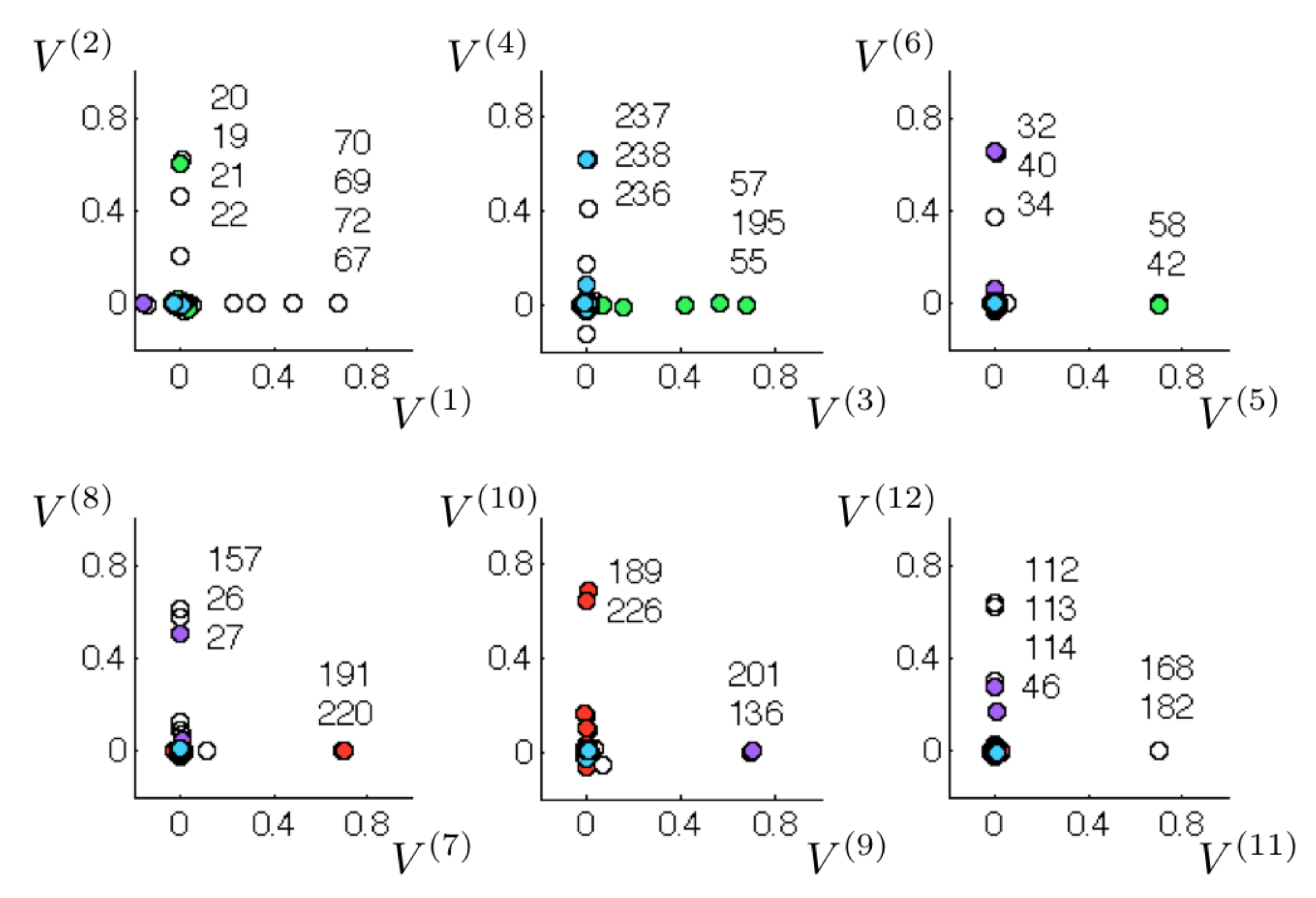}
\caption{Top 24 sectons when not truncating $\mathcal{D}_{ij}$ to $\mathcal{\tilde D}_{ij}$ (except for the strict diagonal, $\mathcal{D}_{ij}=0$) -- Out of the 24 sectons displayed here, 5 consist exclusively of consecutive positions and may be attributed to consecutive gaps.} 
\label{fig:DInotr}
\end{center}
\end{figure}

\clearpage

\begin{table}[t]
\renewcommand{\tablename}{TAB. S}\begin{tabular}{|r|l||r|l||r|l||r|l|}
  \hline
\ Sectons\ &\ Positions\ &\ Sectons\ &\ Positions\ &\ Sectons&\ Positions\ &\ Sectons&\ Positions \\
 \hline
$^*$1 & 224 217  & $^*$31 & 121 45 200  & $^*$61 & 175  & $^*$91 & 96  \\
$^*$2 & 58 42  & $^*$32 & 165 230 176  & 62 & 130 123 166  & $^*$92 & 23  \\
$^*$3 & 57 195 102  & $^*$33 & 114 120 48  & $^*$63 & 28 119  & $^*$93 & 171  \\
$^*$4 & 168 182  & $^*$34 & 235 124 123  & $^*$64 & 110 83  & $^*$94 & 167  \\
$^*$5 & 201 136  & $^*$35 & 67 82 34  & $^*$65 & 117 25  & $^*$95 & 91 101 56 234  \\
$^*$6 & 191 220  & 36 & 197 55 43 19  & 66 & 98 166  & $^*$96 & 115  \\
$^*$7 & 32 40  & $^*$37 & 111 50 49 83  & $^*$67 & 151 143  & $^*$97 & 88  \\
$^*$8 & 226 189  & $^*$38 & 109 84  & $^*$68 & 95  & $^*$98 & 160 228 138  \\
$^*$9 & 44 52 31  & $^*$39 & 27 137  & $^*$69 & 74 153  & 99 & 19 25 33  \\
$^*$10 & 157 26 27  & $^*$40 & 72 78  & $^*$70 & 47 238 53  & $^*$100 & 135 159  \\
$^*$11 & 59 104  & $^*$41 & 214 102 229  & $^*$71 & 148  & $^*$101 & 60  \\
$^*$12 & 194 142  & $^*$42 & 163 225  & $^*$72 & 39  & $^*$102 & 24 70 73  \\
$^*$13 & 30 139 198  & $^*$43 & 144 152  & 73 & 37 124  & $^*$103 & 203  \\
$^*$14 & 112 46 68  & $^*$44 & 221 228 189  & $^*$74 & 22 155  & $^*$104 & 211 228  \\
$^*$15 & 29 122  & $^*$45 & 216 227 190  & $^*$75 & 204  & 105 & 48 125  \\
$^*$16 & 51 105 107  & $^*$46 & 164 132  & $^*$76 & 193 55 43  & $^*$106 & 236  \\
$^*$17 & 209 53 45 121  & 47 & 210 198 162 124 225  & $^*$77 & 75 67  & $^*$107 & 61  \\
$^*$18 & 190 213 138  & 48 & 196 140  & $^*$78 & 173  & $^*$108 & 97  \\
$^*$19 & 184 161  & $^*$49 & 77 70  & $^*$79 & 89  & $^*$109 & 170  \\
$^*$20 & 103 234 229  & $^*$50 & 133 162  & $^*$80 & 54 33  & $^*$110 & 63  \\
$^*$21 & 221A 146  & $^*$51 & 192 143 219  & $^*$81 & 33 41 66  & 111 & 49 113 233  \\
$^*$22 & 107 86 50  & $^*$52 & 108 66 83  & 82 & 188 181 145 221 169  & $^*$112 & 129  \\
$^*$23 & 215 227 180  & $^*$53 & 94 56  & $^*$83 & 64 39  & 113 & 159 137 119 23 176  \\
$^*$24 & 156 21  & $^*$54 & 176 169  & $^*$84 & 31 68 66  & $^*$114 & 147 148  \\
$^*$25 & 179 100 233  & $^*$55 & 231 238 123  & $^*$85 & 87 107  & $^*$115 & 178  \\
$^*$26 & 158 188A 138  & $^*$56 & 93 101  & $^*$86 & 116  & 116 & 34 153  \\
$^*$27 & 199 181 183  & 57 & 128 232 22  & $^*$87 & 184A 221  & 117 & 134 113 62 25  \\
$^*$28 & 118 81 69  & $^*$58 & 92 237  & $^*$88 & 174  & $^*$118 & 200 121  \\
$^*$29 & 85 106 66  & $^*$59 & 177 100  & $^*$89 & 154 21  & $^*$119 & 141 69  \\
$^*$30 & 212 55 45 229  & $^*$60 & 222 187 172  & $^*$90 & 76  & 120 & 125 234  \\
\hline
\end{tabular}
\caption{Composition of sectons -- Sectons are defined here with $k_{\rm top}=120$ and $\epsilon=0.2$. A star indicates that the secton is structurally connected. Up to rank 35, all sectons are thus connected. Note that at large rank, some sectons consist of a single position and are therefore trivially connected; see also Fig.~S\ref{fig:secton_connect}. As indicated in Fig.~S\ref{fig:secton_connect}, structurally disconnected sectons often become connected when considering a more stringent threshold $\epsilon$; this is for instance the case for secton 36 when excluding its last position, 19.}\label{fig:sectons_list}
\end{table}

\clearpage

\begin{figure}[t]
\renewcommand{\figurename}{FIG. S}\begin{center}
\includegraphics[width=.75\linewidth]{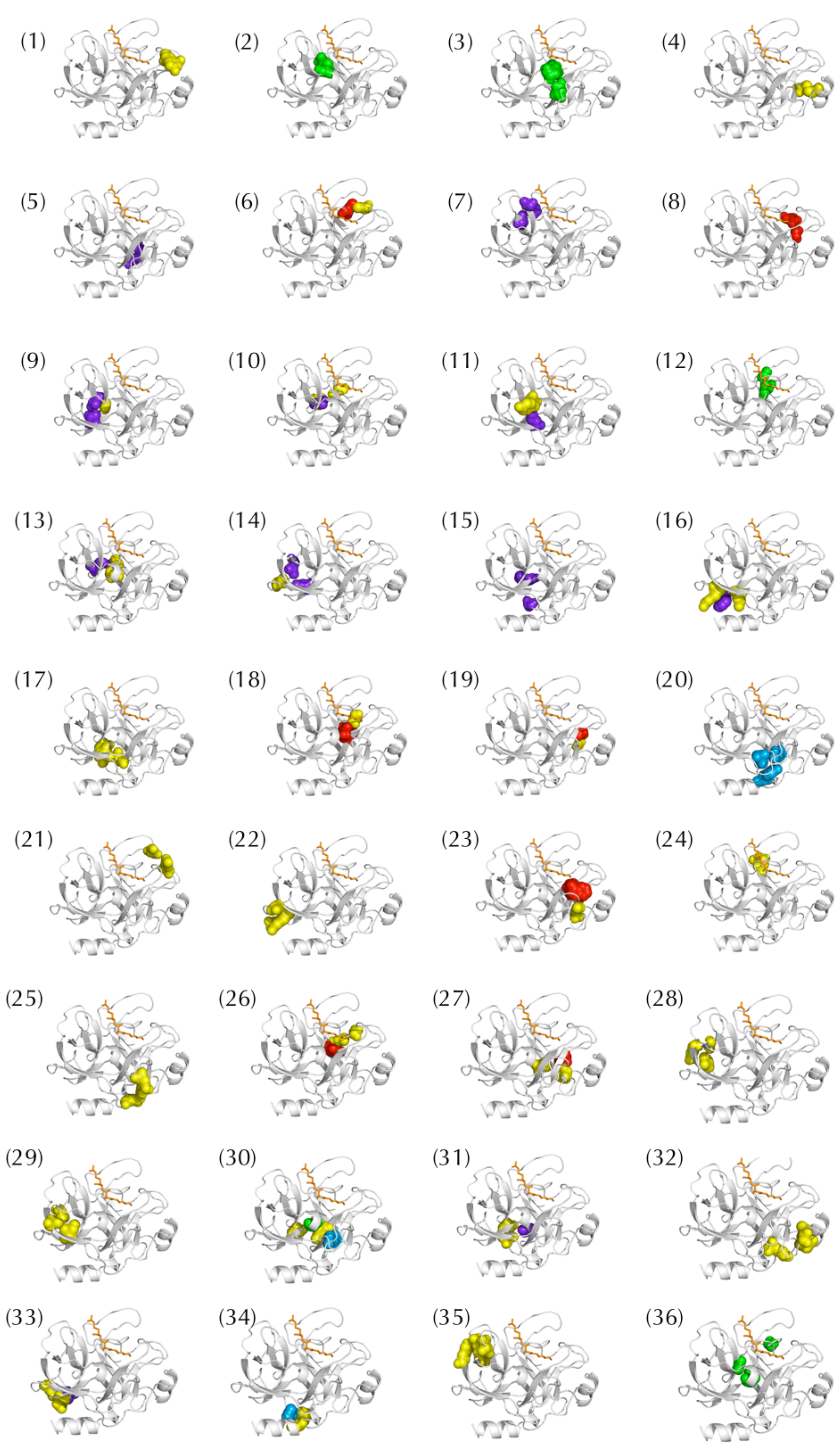}
\caption{Structural representation of the top 36 sectons -- Representation on the three-dimensional structure of rat trypsin of the top 36 sectons defined in Tab.~S\ref{fig:sectons_list}. The colors refer to the sectors, with yellow for non-sector positions (as in Fig.~3, where the first 8 sectons are represented). The 36th secton is the first not to be completely structurally connected, with one position standing apart.}
\label{fig:sectonStruct}
\end{center}
\end{figure}

\begin{figure}[t]
\renewcommand{\figurename}{FIG. S}\begin{center}
\includegraphics[width=.6\linewidth]{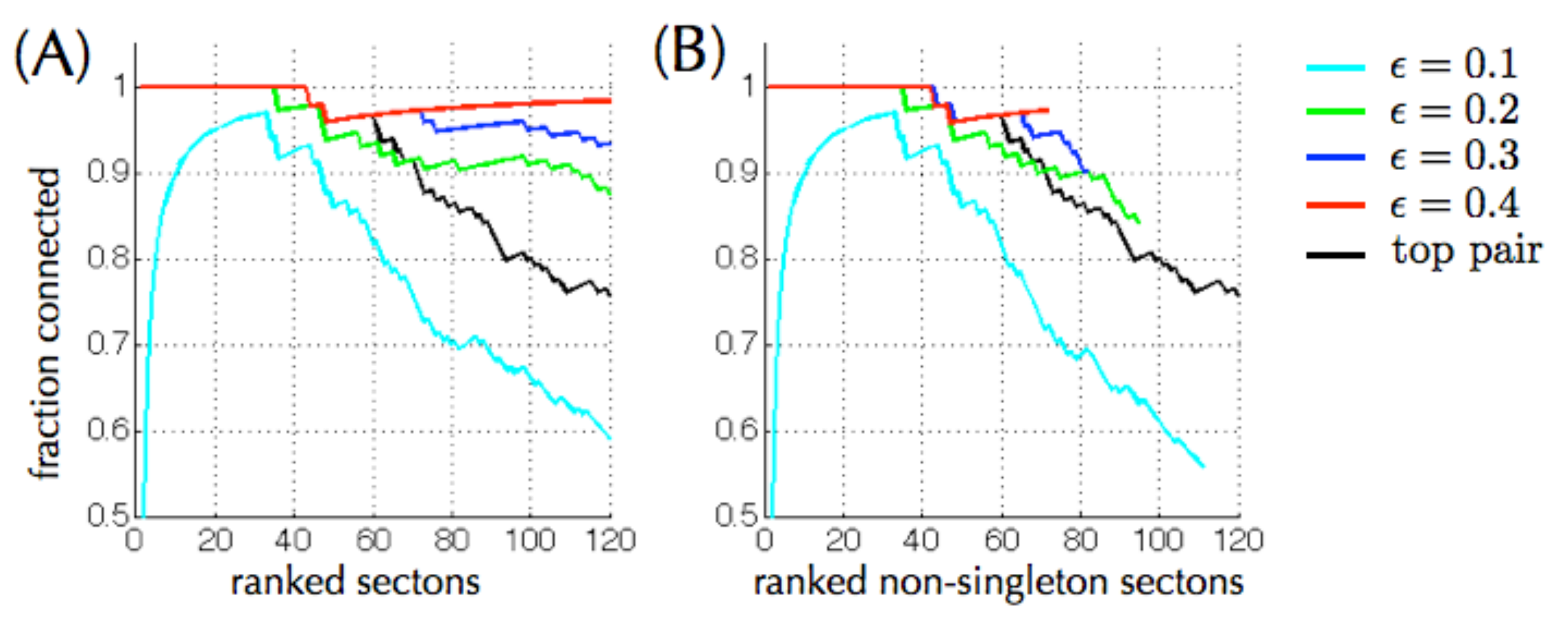}
\caption{Connectivity of sectons -- Sectons are here defined for $k_{\rm top}=120$ and different values of $\epsilon$. A secton is considered as structurally connected if all the positions that it contains are in contact in the three-dimensional structure, either directly or indirectly. {\bf (A)} Fraction of the top sectons to be structurally connected as function of the rank up to which sectons are considered. For comparison, the black curve corresponds to the top 2 positions contributing to $V^{(k)}$. {\bf (B)} Same as (A), but excluding the sectons that contain a single position and are therefore trivially connected. These figures show that by considering a larger, more stringent threshold $\epsilon$ than in Tab.~S\ref{fig:sectons_list} 
(where $\epsilon=0.2$) more (but smaller) sectons are found to be connected.}
\label{fig:secton_connect}
\end{center}
\end{figure}

\begin{figure}[t]
\renewcommand{\figurename}{FIG. S}\begin{center}
\includegraphics[width=.55\linewidth]{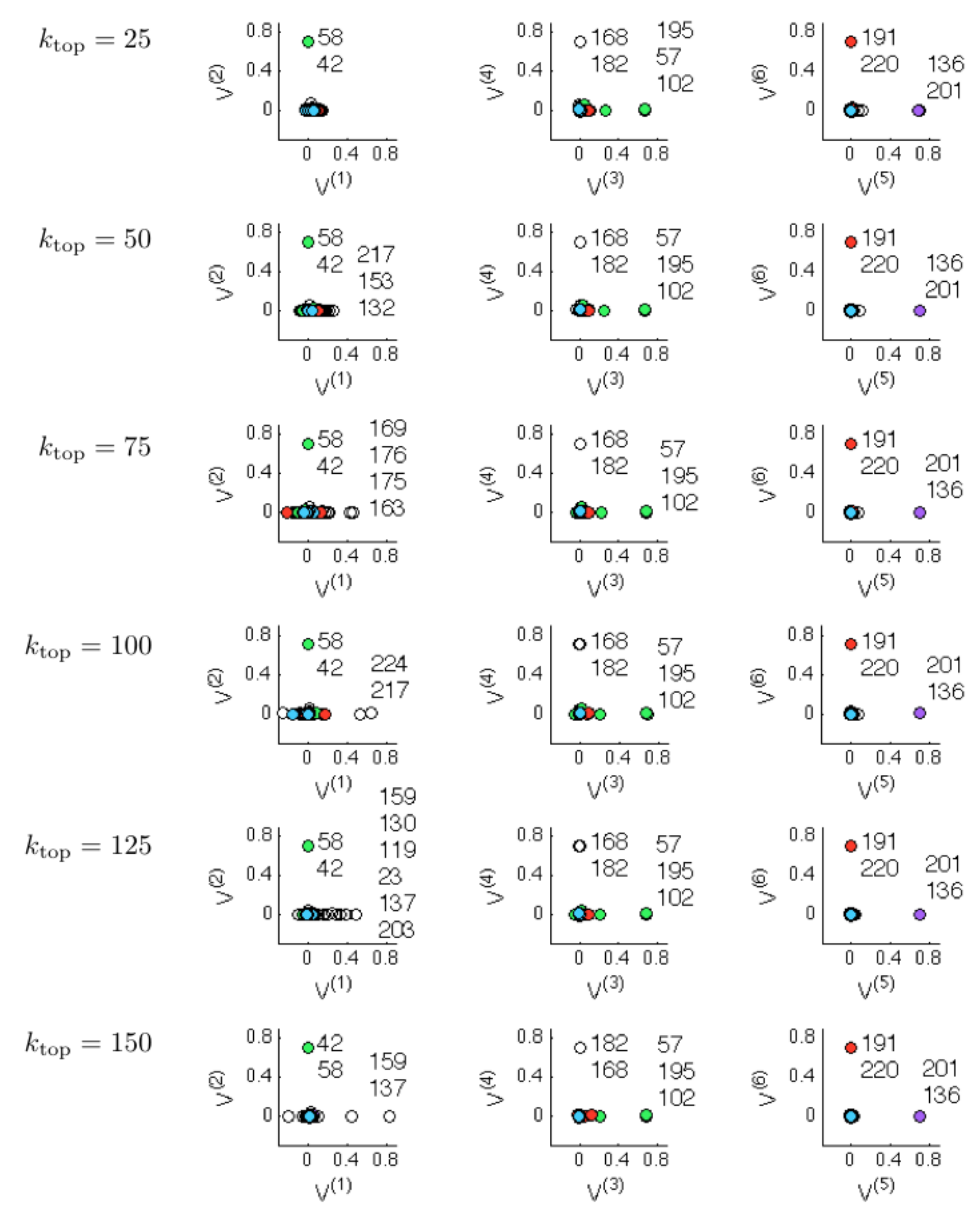}
\caption{Top independent components of $\mathcal{\tilde D}_{ij}$ for varying values of $k_{\rm top}$ -- This figure generalizes Fig.~3, where $k_{\rm top}=120$, to $k_{\rm top}=25,50,75,100,125,150$. With the noticeable exception of the first component $V^{(1)}$, the same components $V^{(k)}$, and therefore the same sectons, are recovered for different values of $k_{\rm top}$. The instability of $V^{(1)}$ suggests that the first secton may better be discarded (although, for $k_{\rm top}=100$ and $k_{\rm top}=150$ at least, it does appear here to convey meaningful information, as the pairs 217-224 and 137-159 are indeed in contact).}
\label{fig:firstsecton}
\end{center}
\end{figure}

\clearpage

\begin{figure}[t]
\renewcommand{\figurename}{FIG. S}\begin{center}
\includegraphics[width=.5\linewidth]{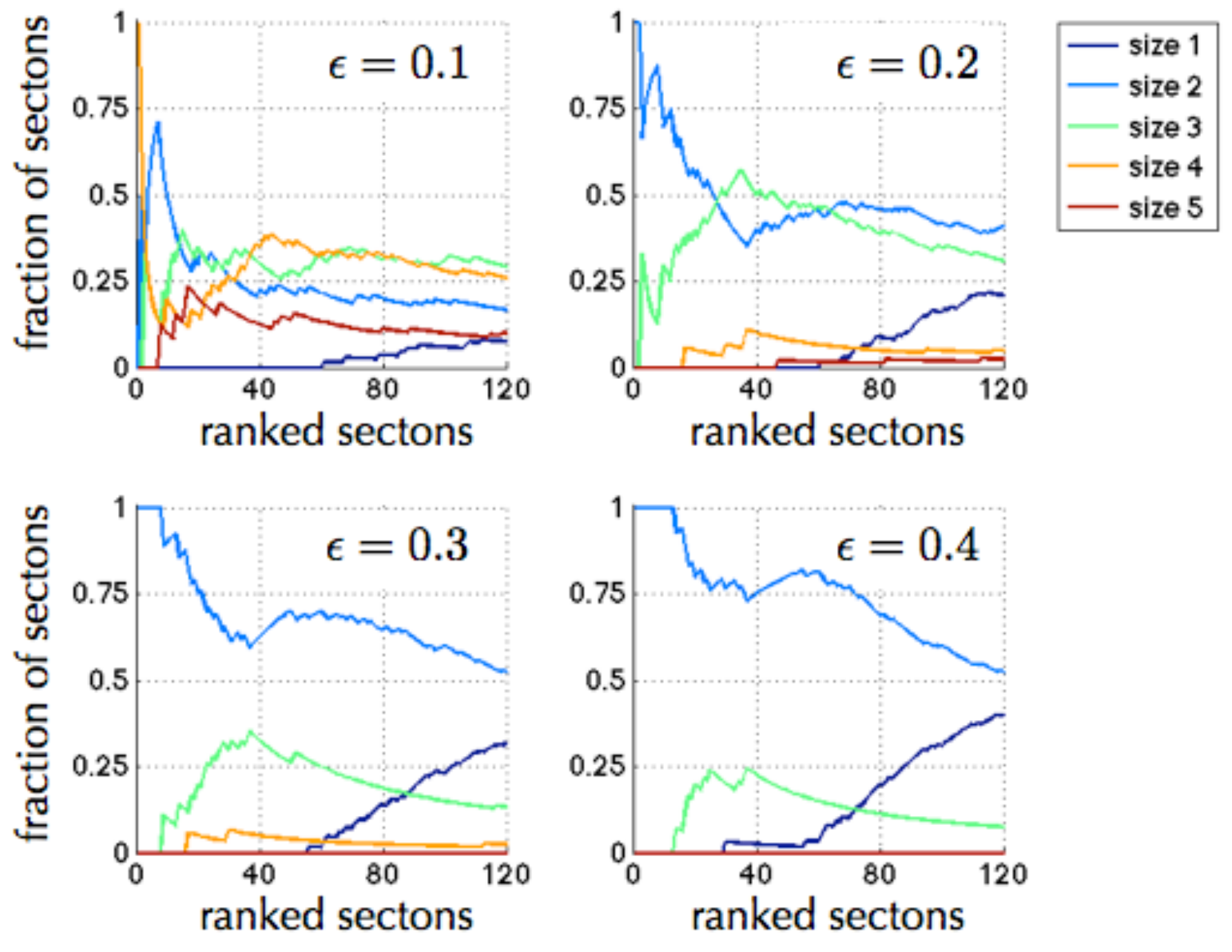}
\caption{Size of sectons -- Distribution of the sizes of sectons, for different values of $\epsilon$. The larger the cut-off $\epsilon$ is, the smaller are the sectons.}
\label{fig:sizeSectons}
\end{center}
\end{figure}

\begin{table}[t]
\renewcommand{\tablename}{TAB. S}\begin{tabular}{|r||l|l|}
  \hline
\ Subgraph\ &\ Positions \ & Secton (if applicable)\\
 \hline
1 &  21 156 & 24 \\
2 &  32 40 & 7 \\
3 &  42 58 & 2 \\
4 &  67 82 & 35$^-$ \\
5 &  81 118 & 28$^-$ \\
6 &  84 109 & 38\\
7 &  85 106 & 29$^-$ \\
8 &  59 104 & 11 \\
9 &  114 120 & 33$^-$ \\
10 &  146 221A & 21 \\
11 &  165 230 & 32$^-$\\
12 &  168 182 & 4 \\
13 &  191 220 & 6 \\
14 &  142 193 194 & 12$^+$\\
15 &  161 184 184A & 19$^+$\\
16 &  180 215 227 & 23 \\
17 &  45 53 121 209 & 17 \\
18 &  100 177 179 233 & \\
19 &  91 101 103 229 234 &  \\
20 &  31 44 46 52 68 108 112 & \\
21 &  50 51 86 87 105 107 111  & \\
22 &  22 26 27 29 30 122 137 139 155 157 200 & \\
23 &  55 57 102 136 138 158 162 181 183 188A & \\
 &  189 190 195 199 201 210 213 214 216 221 226 228 & \\
\hline
\end{tabular}
\caption{Decomposition into connected subgraphs of the contact graph defined from the top pairs of $\mathcal{\tilde D}_{ij}$ -- The top 79 pairs of $\mathcal{\tilde D}_{ij}$, which are all structural contacts, define a contact graph where two pairs are linked if in physical contact (distance $< 8$ \AA). We list here the disjoint connected subgraphs, ranked by size, of this graph, which we compare to sectons. Of the 13 subgraphs of size 2, 7 correspond to sectons of size 2 and 5 to sectons of size 3 (indicated by $^-$). Of the 5 subgraphs of size 3 and 4, 3 correspond to sectons of same size and 2 to sectons of smaller size 3 (indicated by $^+$). Larger subgraphs are also found that generally consist of multiple sectons. This table shows that the sectons are not equivalent to the connected subgraphs of the contact graph defined from the top pairs of $\mathcal{\tilde D}_{ij}$.}\label{fig:subgraphs}
\end{table}

\newpage

\begin{figure}[t]
\renewcommand{\figurename}{FIG. S}\begin{center}
\includegraphics[width=.7\linewidth]{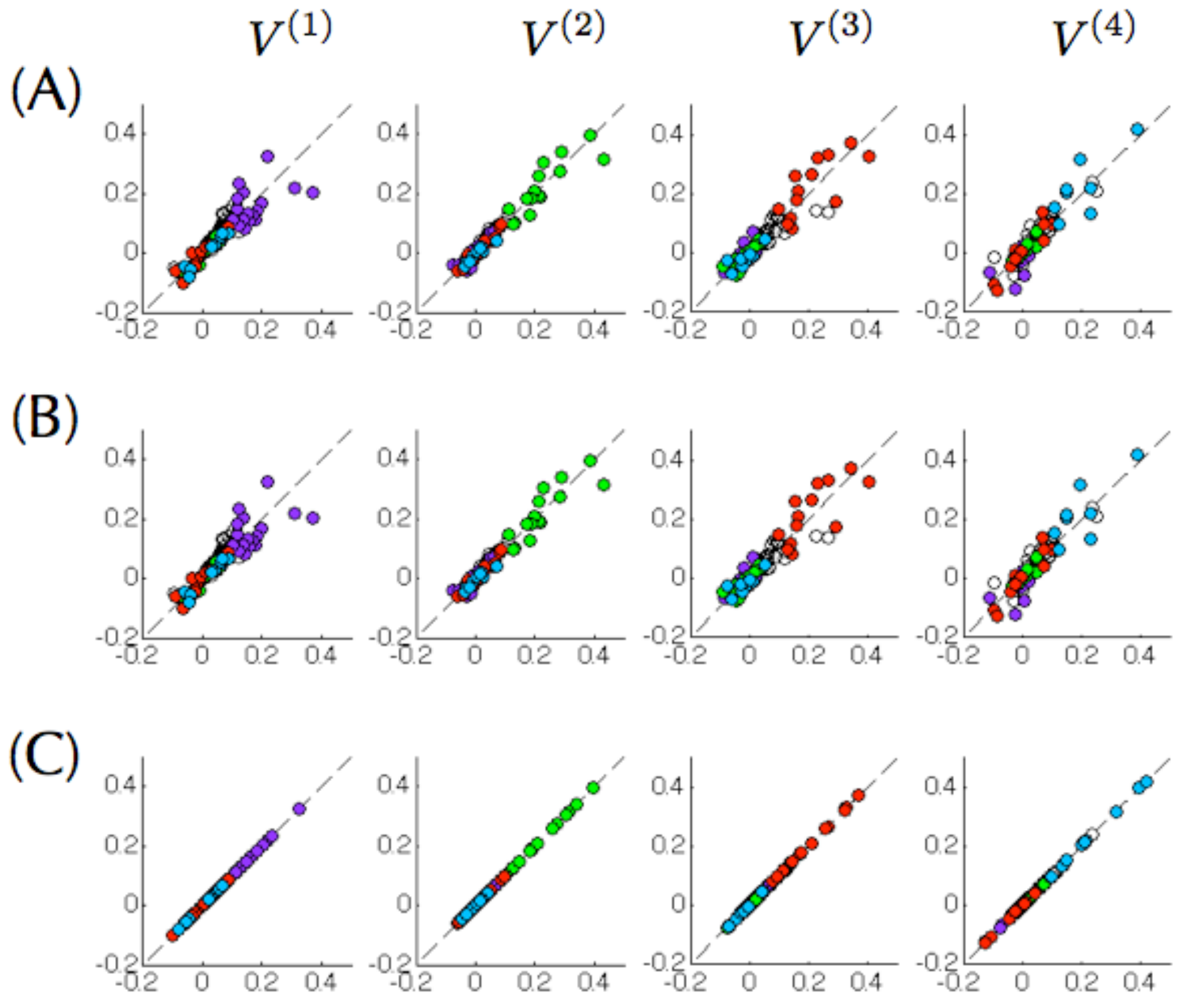}
\caption{ -- Comparison of the top $k_{\rm top}=4$ independent components of SCA matrices defined from the covariance matrix $C$, $\bar C$ and $\bar C^+$ -- ({\bf A}) $V^{(k)}$ from a regularized covariance matrix $\bar C$ ($\mu=1/2$) against $V^{(k)}$ from the original covariance matrix $C$ ($\mu=0$).  Positions are colored as in Fig.~1. ({\bf B}) $V^{(k)}$ from a regularized covariance matrix $\bar C^+$ truncated from its top $k^*$ modes ($\mu=1/2$, $k^*=100$) against $V^{(k)}$ from the original covariance matrix $C$ ($\mu=0$, $k^*=0$). ({\bf C}) $V^{(k)}$ from a regularized covariance matrix $\bar C^+$ truncated from its top $k^*$ modes ($\mu=1/2$, $k^*=100$) against $V^{(k)}$ from the untruncated covariance matrix $C$ ($\mu=1/2$, $k^*=0$). The regularization has a small incidence on the definition of sectors, while the truncation from the top modes has none.}
\label{fig:sectorstop}
\end{center}
\end{figure}

\begin{figure}[t]
\renewcommand{\figurename}{FIG. S}\begin{center}
\includegraphics[width=.7\linewidth]{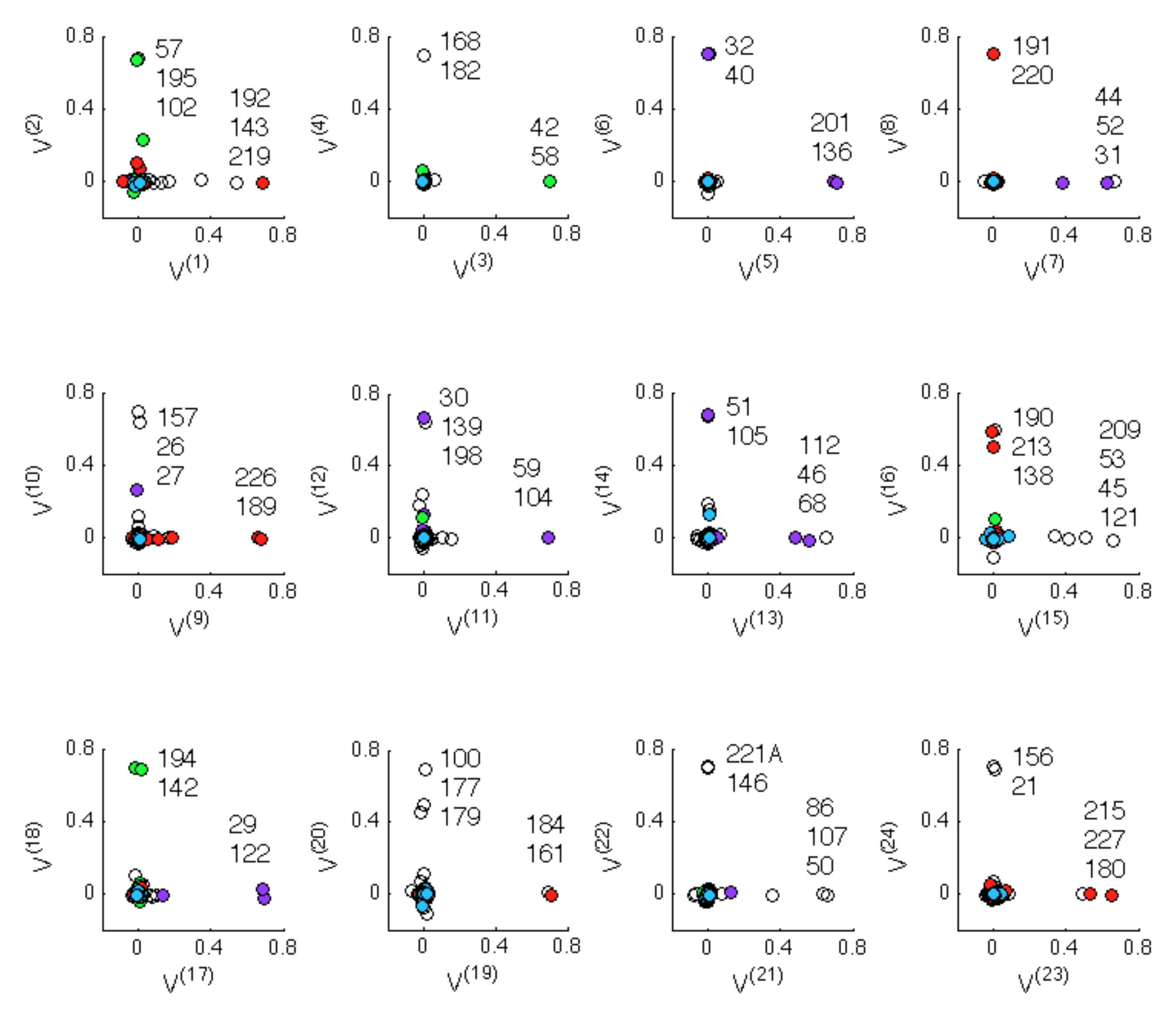}
\caption{Sectons from the bottom modes of $\bar C$ -- We show here the top 24 sectons derived from $\mathcal{\tilde D}_{ij}[J^-]$, with $k^*=100$. A comparison with Fig.~2 shows that essentially the same sectons are defined by this matrix and by $\mathcal{\tilde D}_{ij}[J]$. The exception of the first secton is to be related related to the unstability reported in Fig.~S\ref{fig:firstsecton}.}
\label{fig:sectonsbot}
\end{center}
\end{figure}

\begin{figure}[t]
\renewcommand{\figurename}{FIG. S}\begin{center}
\includegraphics[width=.6\linewidth]{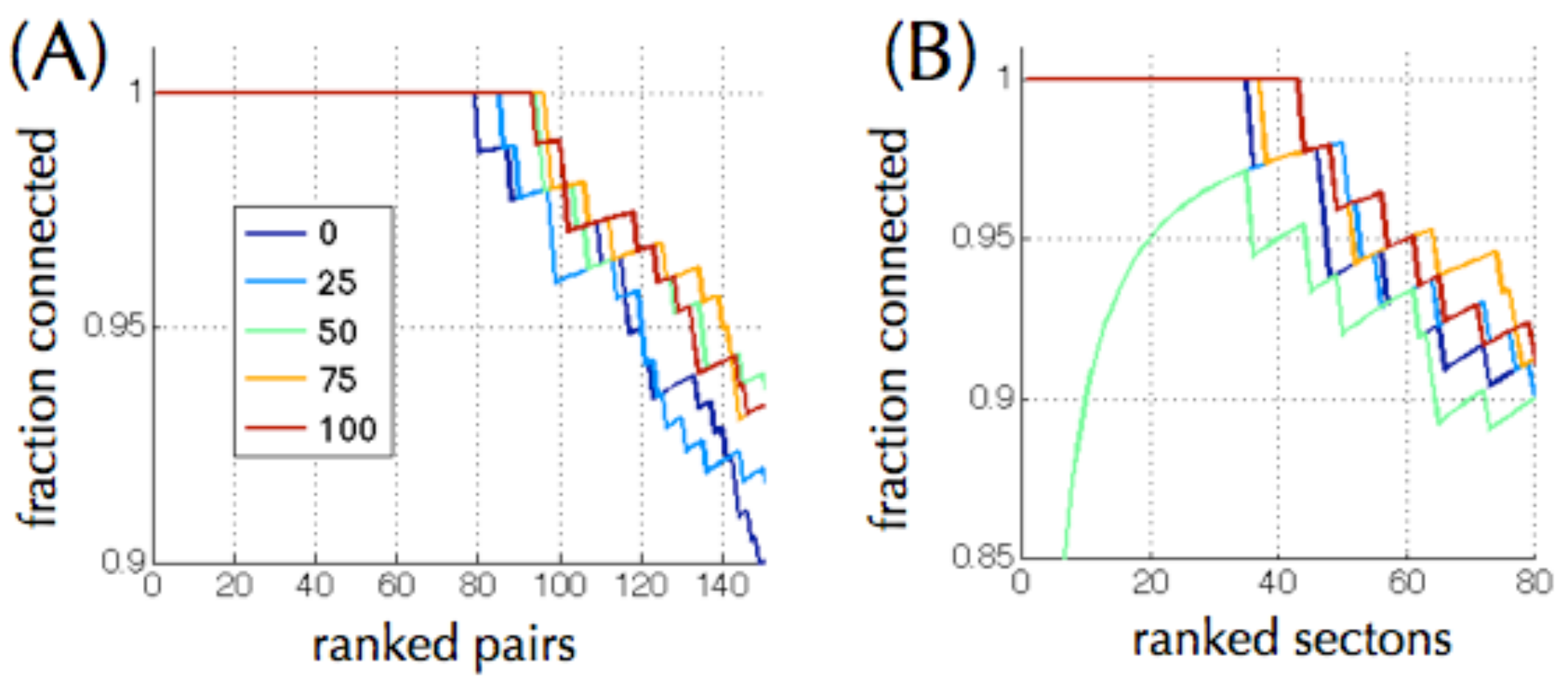}
\caption{Top pairs and sectons from covariance matrices $\bar C$ cleaned from their top modes -- As in the main text, we consider $\bar C=\sum_k|k\rangle\lambda_k\langle k|$, the spectral decomposition of $\bar C$ in the bra-ket notation, with ordered eigenvalues $\lambda_1\geq\dots\geq\lambda_L$. The truncated matrix of direct information $\mathcal{\tilde D}_{ij}$ is defined from
$J^-=-\sum_{k> k^*}|k\rangle\lambda_k^{-1}\langle k|$ for different values of $k^*$, $k^*=0,25,50,75,100$, corresponding to the different curves. Panel (A) for $k^*=0$ (dark blue) corresponds to the results of Fig.~S\ref{fig:truepos} for $\Delta=5$ and panel (B) for $k^*=0$ to Fig.~S\ref{fig:secton_connect}(A) for $\epsilon=0.2$. Truncating the top $k^*=100$ modes of $\bar C$ thus leads to an increase of $\sim 20\%$ in the number of top pairs and top sectons in physical contact. In (B), $k^*=50$ stands apart as a consequence of a disconnected first secton, which, following the observation made in Fig.~S\ref{fig:firstsecton}, could have been ignored (the first 2 positions of this secton are actually connected).}
\label{fig:lessismore}
\end{center}
\end{figure}

\clearpage

\begin{figure}[t]
\renewcommand{\figurename}{FIG. S}
\begin{center}
\includegraphics[width=.7\linewidth]{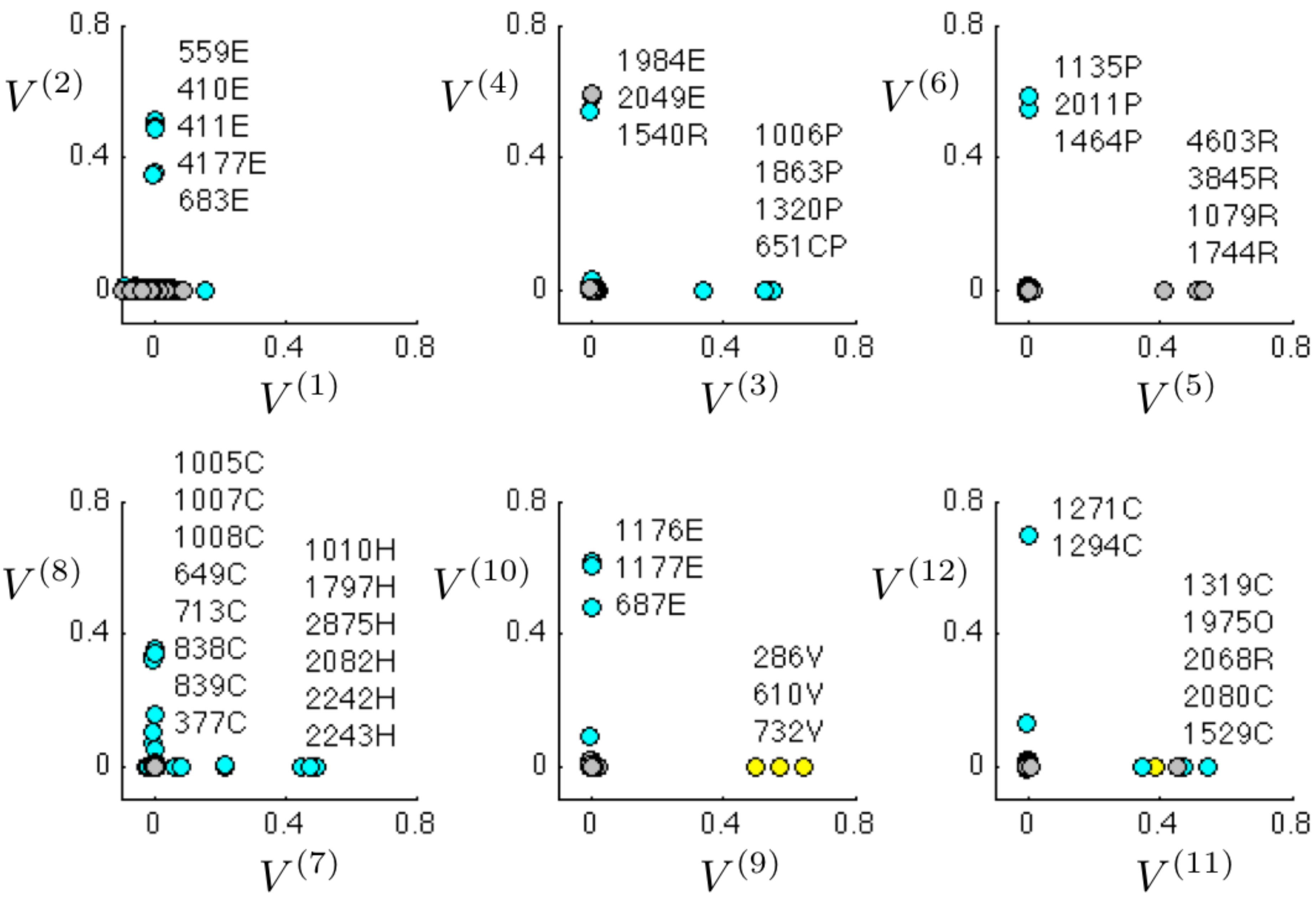}
\caption{Top genomic sectons in bacteria -- As Fig.~3, but for the co-occurrence of orthologous genes in bacterial genomes. Each dot represents a cluster of orthologous genes (COG), with a color associated with its functional class: cyan for metabolism, yellow for cellular processes, magenta for information processing, and gray for poorly characterized genes~\cite{Tatusov:2000tu}. The sectons often comprise genes from a common functional subclass, which is indicated by the last letter labeling the COGs (see Tab.~SV for further genomic sectons).}
\label{fig:gensectons}
\end{center}
\end{figure}

\begin{figure}[t]
\renewcommand{\figurename}{FIG. S}\begin{center}
\includegraphics[width=.6\linewidth]{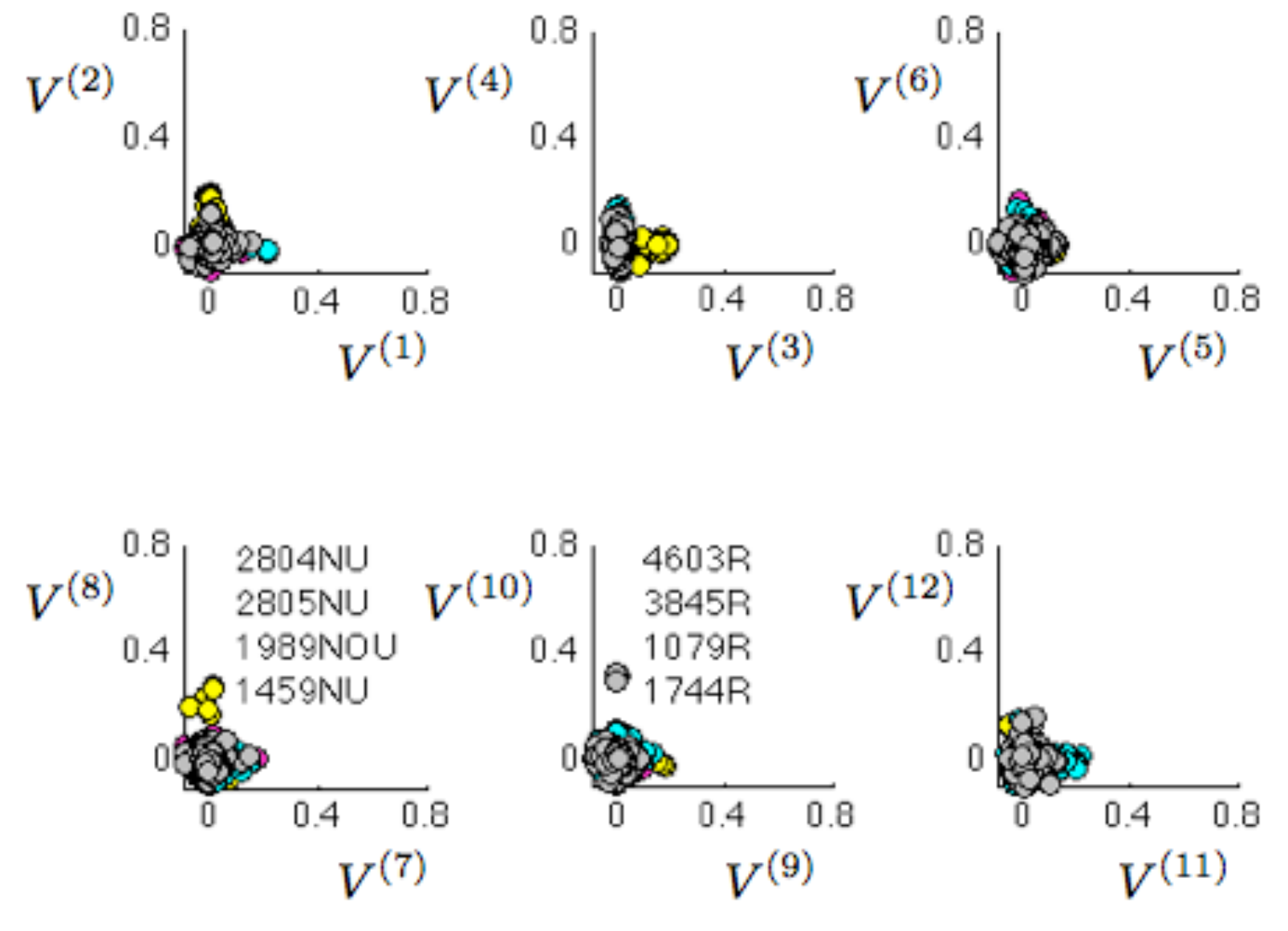}
\caption{Secton analysis of $C_{ij}$ -- The same approach used to define sectons from $\mathcal{D}_{ij}$ is applied to $C_{ij}$. In contrast to Fig.~S\ref{fig:gensectons}, the components are here not localized, thus not defining 'sectons' (truncating the diagonal of $C_{ij}$ does not alter this conclusion).}
\label{fig:SectonC}
\end{center}
\end{figure}

\clearpage

\begin{table}
\renewcommand{\tablename}{TAB. S}
\begin{tabular}{|c|l||c|l|}
\hline
secton & COGs & secton & COGs \\
\hline
1 & & 51 & 1366T 2172T 2208TK \\
2 & 559E 410E 411E 4177E 683E & 52 & 1108P 1121P 803P \\
3 & 1006P 1863P 1320P 651CP & 53 & 5405O 1220O \\
4 & 1984E 2049E 1540R & 54 & 1682GM 1134GM \\
5 & 4603R 3845R 1079R 1744R & 55 & 280C 282C \\
6 & 1135P 2011P 1464P & 56 & 2884D 2177D \\
7 & 1010H 1797H 2875H 2082H 2242H 2243H & 57 & 2224C 2225C \\
8 & 1005C 1007C 1008C 649C 713C 838C 839C 377C & 58 & 1126E 765E 834ET \\
9 & 286V 610V 732V & 59 & 106E 107E 118E 40E 131E 139E 141E \\
10 & 1176E 1177E 687E & 60 & 1663M 1043M 1212M 2877M [...] \\
11 & 1319C 1975O 2068R 2080C 1529C & 61 & 1392P 306P \\
12 & 1271C 1294C & 62 & 588G 696G \\
13 & 163H 43H & 63 & 1354S 1386K \\
14 & 1788I 2057I & 64 & 168P 569P \\
15 & 1346M 1380R & 65 & 1290C 2010C 723C \\
16 & 1653G 3839G 395G 1175G & 66 & 45C 74C \\
17 & 1732M 1125E & 67 & 2801L 2963L \\
18 & 1116P 600P 715P & 68 & 245I 1154HI 1211I 1947I 743I 761IM 821I \\
19 & 1203R 1518L & 69 & 175EH 2895P 7H 155P \\
20 & 1638G 1593G & 70 & 1122P 619P \\
21 & 1640G 58G 296G 297G 1523G 448G & 71 & 1183I 688I \\
22 & 1088M 1091M 1209M 1898M & 72 & 1080G 1762GT 1925G \\
23 & 1129G 1172G 1879G & 73 & 132H 502H 156H 161H \\
24 & 2894D 850D 851D & 74 & 751J 752J \\
25 & 396O 719O & 75 & 1828F 47F \\
26 & 3451U 3505U & 76 & 2009C 1485R 2142C \\
27 & 2332O 2386O 1138O & 77 & 263E 14E \\
28 & 1270H 2038H 2087H 1492H 368H & 78 & 4799I 511I 777I 825I \\
29 & 1677NU 1684NU 1766NU 1157NU [...] & 79 & 1180O 1328F \\
30 & 22C 1071C 508C & 80 & 1013C 674C \\
31 & 1228Q 2986E 2987E & 81 & 113H 181H 1H 373H \\
32 & 1120PH 609P 614P & 82 & 3275T 3279KT \\
33 & 1622C 109O 1845C 843C 1612O & 83 & 413H 414H 853H \\
34 & 3288C 1282C & 84 & 547E 133E 134E 135E 147EH 512EH 159E \\
35 & 1918P 370P & 85 & 1704S 1512R \\
36 & 2804NU 1989NOU 1450NU 1459NU & 86 & 554C 578C \\
37 & 117H 1985H & 87 & 602O 603R 720H \\
38 & 2896H 303H 746H 521H 314H 315H & 88 & 1077D 1792M 1426S \\
39 & 4962U 4965U & 89 & 4149P 725P \\
40 & 241E 279G 2870M 859M & 90 & 3383R 437C 1526C \\
41 & 1003E 404E 509E & 91 & 1020Q 2091H 736I \\
42 & 2025C 2086C & 92 & 1706N 1261NO 2063N \\
43 & 422H 2022H 2104H & 93 & 55C 56C 355C 356C 711C 712C 224C \\
44 & 1173EP 601EP 747E 444EP & 94 & 473CE 119E 129EG 59EH 65E 66E 440E \\
45 & 226P 1117P 573P 581P & 95 & 1837R 1847R \\
46 & 379H 29H 157H & 96 & 363G 364G \\
47 & 1352NT 643NT 2201NT 835NT 840NT & 97 & 2805NU 1989NOU 4972NU \\
48 & 1178P 1840P & 98 & 548E 1364E 4992E 2E \\
49 & 1034C 1894C 1905C & 99 & 419L 420L \\
50 & 1127Q 1463Q 767Q & 100 & 455D 1191K 1317NU 1345N 1419N 1516NUO 3144N \\
\hline
\end{tabular}
\caption{Content of the top 100 genomic sectons in terms of COGs --  The full content of secton 29 is $\{$1677NU 1684NU 1766NU 1157NU 1815N 1843N 1868N 1256N 1291N 1298NU 1338NU 1344N 1987NU 1377NU 1536N 1558N$\}$ and the full content of secton 60 is $\{$1663M 1043M 1212M 2877M 763M 774M 1519M 794M 1560M$\}$.} \label{fig:gensectons_list}   
\end{table}

\clearpage

\begin{figure}[t]
\renewcommand{\figurename}{FIG. S}\begin{center}
\includegraphics[width=.5\linewidth]{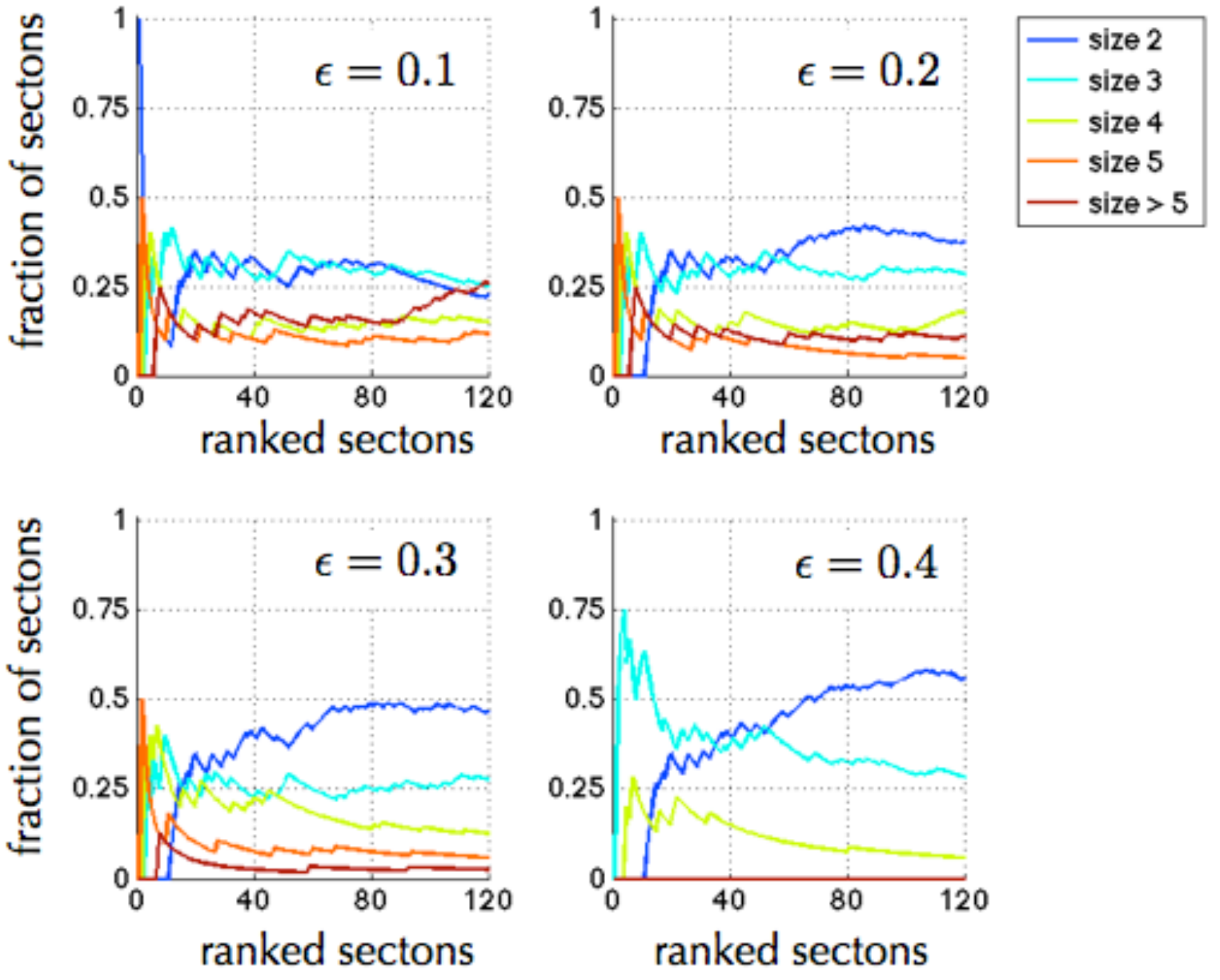}
\caption{Size of genomic sectons -- Distribution of the sizes of sectons, for different values of $\epsilon$. The larger the cut-off $\epsilon$ is, the smaller are the sectons (sectons of size 1 are in negligible number).}
\label{fig:sizeSectGen}
\end{center}
\end{figure}

\begin{figure}[t]
\renewcommand{\figurename}{FIG. S}\begin{center}
\includegraphics[width=.45\linewidth]{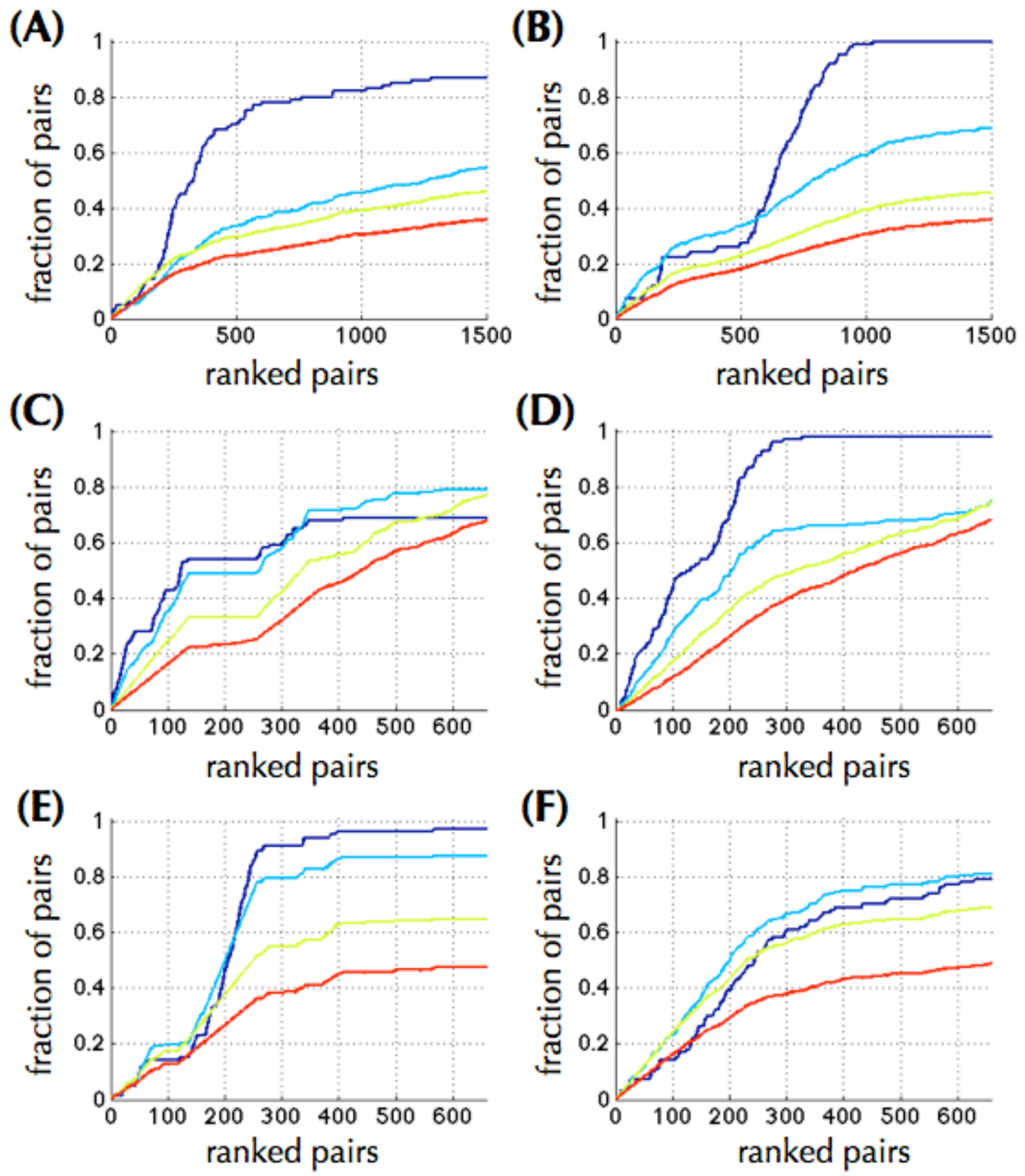}
\caption{Comparison of top pairs for $C_{ij}$, $\mathcal{D}_{ij}$ and sectons -- {\bf (A)} To compare the top pairs for $C_{ij}$ and $\mathcal{D}_{ij}$, we represent the fraction of top pairs for $C_{ij}$, up to the rank indicated along the $x$-axis,  to be included in the top 100 (blue curve), 500 (cyan), 1000 (green) or 1500 (red) top pairs for $\mathcal{D}_{ij}$. This shows that $\sim 40\%$ of the top $n$ pairs for $C_{ij}$ are in the top $n$ pairs for $C_{ij}$. {\bf (B)} Same as (A) but exchanging the roles of $C_{ij}$ and $\mathcal{D}_{ij}$. {\bf (C)} To compare with the content of sectons, we listed the pairs contained in sectons, ranked them by rank of sectons (with arbitrary ranking within sectons) and represented the fraction of these top pairs, up to the rank indicated along the $x$-axis,  to be included in the top 100 (blue curve), 200 (cyan), 400 (green) or 600 (red) top pairs for $\mathcal{D}_{ij}$. {\bf (D)} Same as (C) but exchanging the roles of top pairs for sectons and top pairs for $\mathcal{D}_{ij}$. {\bf (E)} Same as (C) but for $C_{ij}$ instead of $\mathcal{D}_{ij}$. {\bf (F)} Same as (D) but for $C_{ij}$ instead of $\mathcal{D}_{ij}$. Overall, the last 4 panels show an overlap of $\sim 60\%$ between pairs in top sectons and top pairs for $\mathcal{D}_{ij}$ and $\sim 40\%$ between pairs in top sectons and top pairs for $C_{ij}$.}
\label{fig:ComparCDS}
\end{center}
\end{figure}

\clearpage

\begin{table}
\renewcommand{\tablename}{TAB. S}
\begin{tabular}{|c|l|l|}
\hline
Secton & COG & Annotation\\
\hline
\hline
2 & 559E & Branched-chain amino acid ABC-type transport system, permease components \\
 & 410E & ABC-type branched-chain amino acid transport systems, ATPase component \\
 & 411E & ABC-type branched-chain amino acid transport systems, ATPase component \\
 & 4177E & ABC-type branched-chain amino acid transport system, permease component \\
 & 683E & ABC-type branched-chain amino acid transport systems, periplasmic component \\
\hline
3 & 1006P & Multisubunit Na+/H+ antiporter, MnhC subunit \\
 & 1863P & Multisubunit Na+/H+ antiporter, MnhE subunit \\
 & 1320P & Multisubunit Na+/H+ antiporter, MnhG subunit \\
 & 651CP & Formate hydrogenlyase subunit 3/Multisubunit Na+/H+ antiporter, MnhD subunit \\
\hline
4 & 1984E & Allophanate hydrolase subunit 2 \\
 & 2049E & Allophanate hydrolase subunit 1 \\
 & 1540R & Uncharacterized proteins, homologs of lactam utilization protein B \\
\hline
5 & 4603R & ABC-type uncharacterized transport system, permease component \\
 & 3845R & ABC-type uncharacterized transport systems, ATPase components \\
 & 1079R & Uncharacterized ABC-type transport system, permease component \\
 & 1744R & Uncharacterized ABC-type transport system, periplasmic component/surface lipoprotein \\
\hline
6 & 1135P & ABC-type metal ion transport system, ATPase component \\
 & 2011P & ABC-type metal ion transport system, permease component \\
 & 1464P & ABC-type metal ion transport system, periplasmic component/surface antigen \\
\hline
7 & 1010H & Precorrin-3B methylase \\
 & 1797H & Cobyrinic acid a,c-diamide synthase \\
 & 2875H & Precorrin-4 methylase \\
 & 2082H & Precorrin isomerase \\
 & 2242H & Precorrin-6B methylase 2 \\
 & 2243H & Precorrin-2 methylase \\
\hline
8 & 1005C & NADH:ubiquinone oxidoreductase subunit 1 (chain H) \\
 & 1007C & NADH:ubiquinone oxidoreductase subunit 2 (chain N) \\
 & 1008C & NADH:ubiquinone oxidoreductase subunit 4 (chain M) \\
 & 649C & NADH:ubiquinone oxidoreductase 49 kD subunit 7 \\
 & 713C & NADH:ubiquinone oxidoreductase subunit 11 or 4L (chain K) \\
 & 838C & NADH:ubiquinone oxidoreductase subunit 3 (chain A) \\
 & 839C & NADH:ubiquinone oxidoreductase subunit 6 (chain J) \\
 & 377C & NADH:ubiquinone oxidoreductase 20 kD subunit and related Fe-S oxidoreductases \\
\hline
9 & 286V & Type I restriction-modification system methyltransferase subunit \\
 & 610V & Type I site-specific restriction-modification system, R (restriction) subunit and related helicases \\
 & 732V & Restriction endonuclease S subunits \\
\hline
10 & 1176E & ABC-type spermidine/putrescine transport system, permease component I \\
 & 1177E & ABC-type spermidine/putrescine transport system, permease component II \\
 & 687E & Spermidine/putrescine-binding periplasmic protein \\
\hline
11 & 1319C & Aerobic-type carbon monoxide dehydrogenase, middle subunit CoxM/CutM homologs \\
 & 1975O & Xanthine and CO dehydrogenases maturation factor, XdhC/CoxF family \\
 & 2068R & Uncharacterized MobA-related protein \\
 & 2080C & Aerobic-type carbon monoxide dehydrogenase, small subunit CoxS/CutS homologs \\
 & 1529C & Aerobic-type carbon monoxide dehydrogenase, large subunit CoxL/CutL homologs \\
\hline
12 & 1271C & Cytochrome bd-type quinol oxidase, subunit 1 \\
 & 1294C & Cytochrome bd-type quinol oxidase, subunit 2 \\
\hline
\end{tabular}
\caption{Annotations for the top genomic sectons -- Note that there is no secton along the first component, as seen in Fig.~S\ref{fig:gensectons}. (Sectons 13 to 24 are presented on the next page). This shows that the definition of sectons is consistent with our current knowledge of gene functions with, in particular, many of the top sectons involving different subunits of a same protein complex.}\label{tab:sectons2}
\end{table}

\clearpage

\begin{figure}
\renewcommand{\figurename}{FIG. S}
\begin{tabular}{|c|l|l|}
\hline
11 & 1319C & Aerobic-type carbon monoxide dehydrogenase, middle subunit CoxM/CutM homologs \\
 & 1975O & Xanthine and CO dehydrogenases maturation factor, XdhC/CoxF family \\
 & 2068R & Uncharacterized MobA-related protein \\
 & 2080C & Aerobic-type carbon monoxide dehydrogenase, small subunit CoxS/CutS homologs \\
 & 1529C & Aerobic-type carbon monoxide dehydrogenase, large subunit CoxL/CutL homologs \\
\hline
12 & 1271C & Cytochrome bd-type quinol oxidase, subunit 1 \\
 & 1294C & Cytochrome bd-type quinol oxidase, subunit 2 \\
\hline
13 & 163H & 3-polyprenyl-4-hydroxybenzoate decarboxylase \\
 & 43H & 3-polyprenyl-4-hydroxybenzoate decarboxylase and related decarboxylases \\
\hline
14 & 1788I & Acyl CoA:acetate/3-ketoacid CoA transferase, alpha subunit \\
 & 2057I & Acyl CoA:acetate/3-ketoacid CoA transferase, beta subunit \\
\hline
15 & 1346M & Putative effector of murein hydrolase \\
 & 1380R & Putative effector of murein hydrolase LrgA \\
\hline
16 & 1653G & ABC-type sugar transport system, periplasmic component \\
 & 3839G & ABC-type sugar transport systems, ATPase components \\
 & 395G & ABC-type sugar transport system, permease component \\
 & 1175G & ABC-type sugar transport systems, permease components \\
\hline
17 & 1732M & Periplasmic glycine betaine/choline-binding (lipo)protein of an ABC-type transport system \\
 & 1125E & ABC-type proline/glycine betaine transport systems, ATPase components \\
\hline
18 & 1116P & ABC-type nitrate/sulfonate/bicarbonate transport system, ATPase component \\
 & 600P & ABC-type nitrate/sulfonate/bicarbonate transport system, permease component \\
 & 715P & ABC-type nitrate/sulfonate/bicarbonate transport systems, periplasmic components \\
\hline
19 & 1203R & Predicted helicases \\
 & 1518L & Uncharacterized protein predicted to be involved in DNA repair \\
\hline
20 & 1638G & TRAP-type C4-dicarboxylate transport system, periplasmic component \\
 & 1593G & TRAP-type C4-dicarboxylate transport system, large permease component \\
\hline
21 & 1640G & 4-alpha-glucanotransferase \\
 & 58G & Glucan phosphorylase \\
 & 296G & 1,4-alpha-glucan branching enzyme \\
 & 297G & Glycogen synthase \\
 & 1523G & Type II secretory pathway, pullulanase PulA and related glycosidases \\
 & 448G & ADP-glucose pyrophosphorylase \\
\hline
22 & 1088M & dTDP-D-glucose 4,6-dehydratase \\
 & 1091M & dTDP-4-dehydrorhamnose reductase \\
 & 1209M & dTDP-glucose pyrophosphorylase \\
 & 1898M & dTDP-4-dehydrorhamnose 3,5-epimerase and related enzymes \\
\hline
23 & 1129G & ABC-type sugar transport system, ATPase component \\
 & 1172G & Ribose/xylose/arabinose/galactoside ABC-type transport systems, permease components \\
 & 1879G & ABC-type sugar transport system, periplasmic component \\
\hline
24 & 2894D & Septum formation inhibitor-activating ATPase \\
 & 850D & Septum formation inhibitor \\
 & 851D & Septum formation topological specificity factor \\
\hline
\end{tabular}
\end{figure}

\clearpage

\begin{figure}[t]
\renewcommand{\figurename}{FIG. S}\begin{center}
\includegraphics[width=.9\linewidth]{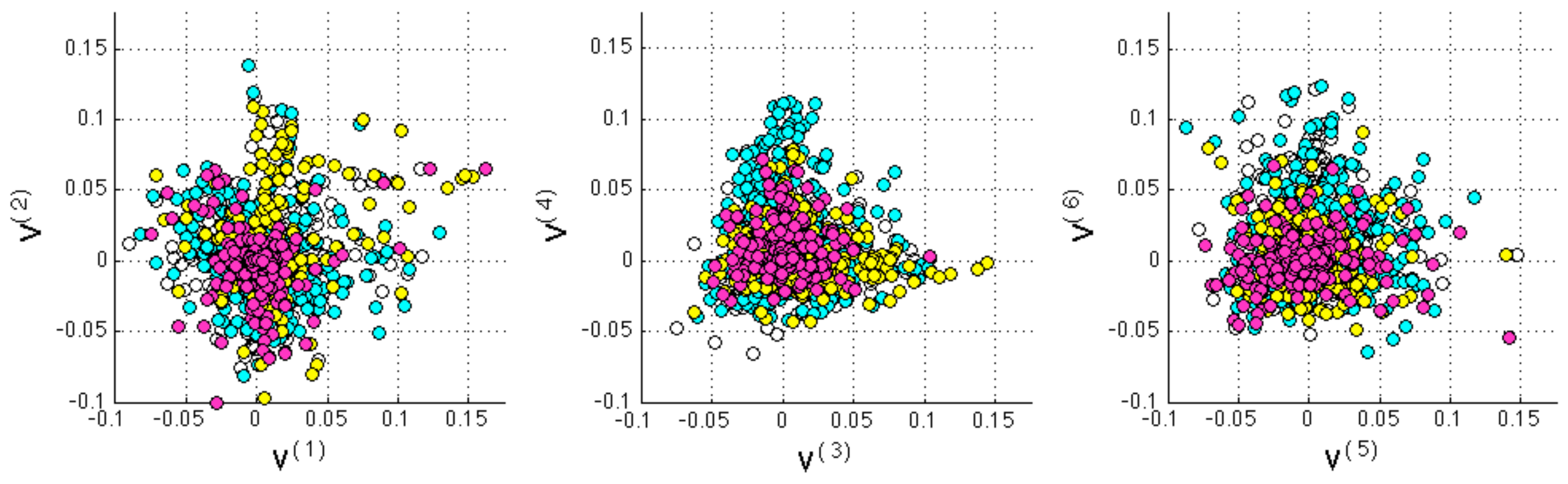}
\caption{Genomic sectors ($k_{\rm max}=6$) -- COGs are colored as in Fig.~S\ref{fig:gensectons} according to the functional category to which they belong: cyan for metabolism, yellow for cellular processes, magenta for information processing, and white for poorly characterized genes or genes that belong to multiple categories~\cite{Tatusov:2000tu}. The apparent enrichment of yellow COGs along $V^{(2)}$ or cyan COGs along $V^{(4)}$ is quantitatively estimated in Tab.~S\ref{fig:SectorsGenStat}.}
\label{fig:SectorsGen}
\end{center}
\end{figure}

\begin{table}[t]
\renewcommand{\tablename}{TAB. S}
\begin{tabular}{|c|c|c|c|c|c|}
  \hline
\ Sector\ &\ Total number of genes\ &\ Info. process.\ &\ Cellular process.\ &\ Metabolism\ &\ $p$-value for the composition\\
 \hline
1 & 22 & 1 & 4 & 13 & 0.14 \\
2 & 56 & 5 & 25 & 17 & 5 $10^{-5}$ \\
3 & 74 & 2 & 35 & 23 & 5 $10^{-9}$ \\
4 & 114 & 8 & 10 & 85 & 3 $10^{-9}$\\
5 & 50 & 12 & 7 & 24 & 0.37 \\
6 & 65 & 2 & 4 & 32 & 3 $10^{-4}$ \\\hline
\end{tabular}
\caption{Association between genomic sectors and functional categories -- For each sector, defined as the COGs with contribution $>0.05$ along one of the 6 components $V^{(k)}$ shown in Fig.~S\ref{fig:SectorsGen} and not along any other, we assessed the significance of their content in information processing (magenta), cellular processing (yellow), metabolic (cyan) COGs , which represent respectively around $1/2$, $1/4$ and $1/4$ of the COGs. The $p$-value is computed from a $\chi^2$-square test with 2 degrees of freedom. 4 of the 6 sectors can be considered to be significantly enriched in COGs of some of these functional categories.}\label{fig:SectorsGenStat}
\end{table}

\end{document}